Modern Education
and Computer Science
PRESS

# An Extended Symbolic-Arithmetic Model for Teaching Double-Black Removal with Rotation in Red-Black Trees

**Kennedy E. Ehimwenma\***
Department of Computer Science College of Science, Mathematics and Technology, Wenzhou Kean University, China
Email: kehimwen@kean.edu
ORCID iD: https://orcid.org/0000-0002-7616-9342
\*Corresponding Author

**Hongyu Zhou**
Department of Computer Science College of Science, Mathematics and Technology, Wenzhou Kean University, China
Email: zhouho@kean.edu
ORCID iD: https://orcid.org/0009-0002-8762-0378

**Junfeng Wang**
Department of Computer Science College of Science, Mathematics and Technology, Wenzhou Kean University, China
Email: wangjunf@kean.edu
ORCID iD: https://orcid.org/0009-0003-7897-1480

**Ze Zheng**
Department of Computer Science College of Science, Mathematics and Technology, Wenzhou Kean University, China
Email: zhenze@kean.edu
ORCID iD: https://orcid.org/0009-0009-8222-7826

**Abstract:** Double-black (DB) nodes have no place in red-black (RB) trees. So when DB nodes are formed, they are immediately removed. The removal of DB nodes that cause rotation and recoloring of other connected nodes poses greater challenges in the teaching and learning of RB trees. To ease this difficulty, this paper extends our previous work on the symbolic arithmetic algebraic (SA) method for removing DB nodes. The SA operations that are given as, *Red + Black = Black; Black - Black = Red; Black + Black = DB;* and *DB − Black = Black* removes DB nodes and rebalances black heights in RB trees. By extension, this paper projects three SA mathematical equations, namely, *general symbolic arithmetic rule,* $\Delta_{[DB,r,p]}$; *partial symbolic arithmetic rule1,* $\partial'_{[DB,p]}$; and *partial symbolic arithmetic rule2,* $\partial''_{[r]}$. The removal of a DB node ultimately affects black heights in RB trees. To balance black heights using the SA equations, all the RB tree cases, namely, LR, RL, LL, and RR, were considered in this work; and the position of the nodes connected directly or indirectly to the DB node was also tested. In this study, to balance a RB tree, the issues considered *w.r.t.* the different cases of the RB tree were i) whether a DB node has an inner, outer, or both inner and outer black nephews; or ii) ) whether a DB node has an inner, outer or both inner and outer red nephews. The *nephews r and x* in this work are the children of the *sibling s* to a DB, and further up the tree, the *parent p* of a DB is their *grandparent g*. Thus, *r* and *x* have indirect relationships to a DB at the point of formation of the DB node. The novelty of the SA equations is in their effectiveness in the removal of DB that involves rotation of nodes as well as the recoloring of nodes along any simple path so as to balance *black heights* in a tree. Our SA methods assert when, where, and how to remove a DB node and the nodes to recolor. As shown in this work, the SA algorithms have been demonstrated to be faster in performance *w.r.t.* to the number of steps taken to balance a RB tree when compared to the traditional RB algorithm for DB removal. The simplified and systematic approach of the SA methods has enhanced student learning and understanding of DB node removal in RB trees.

## 1. Introduction

Red black (RB) trees are binary trees and by extension binary search trees (BST) [1, 2, 3]. The color of the nodes in a RB can be either red or black. Any node, whether red or black, can be inserted or deleted as result of data update or some modification in the RB tree data structure [4]. Modifications in a RB tree can be from the insertion or deletion on of node. The resultant color of nodes in a balanced tree is a reflection of such modifications. In RB trees, the removal of **DB** nodes follows some procedure and systematic stages that are very complex. The complexity that lies in the removal of **DB** node which involves some node rotation in the tree and color change are relatively fluid and ambiguous during classroom teaching. To resolve these ambiguities, the questions to ask are *i) How is a DB node formed? ii) How can a DB be removed? iii) What is the principle behind the formation of DB? iv) At what point should a DB be removed if we recur up the RB tree? v) Which node(s) are involved in color change? and vi) What are principles governing the recoloring of nodes?* The foregoing forms part of the many reasons why students encounter difficulty in understanding the principles that surrounds the removal of **DB** nodes. The lack of absolute clarity in the procedure for **DB** removal in literature is what this paper has addressed using formal logic and symbolic mathematical procedures. This mathematical procedure is an extension of our initial work in which we state that, *Red + Black = Black; Black - Black = Red; Black + Black = DB; Black + NULLLEAF = DB and DB - Black = Black.*

The goal of this work is to project a model that can ease the learning of **DB** node removal in RB trees. This paper thus proposes three different equations called the *symbolic arithmetic* rules. These equations, namely, *general symbolic arithmetic rule (GSAR)* given as $\Delta_{[DB,r,p]}$; *partial symbolic arithmetic rule1 (PSAR1)* as $\partial'_{[DB,p]}$; and *partial symbolic arithmetic rule2 (PSAR2)* as $\partial''_{[r]}$ mathematically computes the stages involved in **DB** removal and recoloring of the affected nodes along any simple path. The stages demystified by the mathematical rules demonstrate to students the step-by-step process of removing a **DB** node and recoloring the affected nodes. To the best of our knowledge, this SA approach is novel in the use of symbolic-arithmetic equations in DB node removal, recoloring of nodes, and rebalancing of black heights in a RB tree.

### 1.1 Node Color-Change

Our recent work in [4] projected and demonstrated the SA method for the balancing of RB trees after a delete operation is **performed** on a node. Ehimwenma *et al.* [4] states that the deletion of a red (leaf) node and the rebalancing of a RB tree is pretty straightforward. That; with the symbolic expression; ***Red + Black = Black***, and ***Red + NULL_LEAF = Black***; any deleted red color node is replaced by a black color node. Similarly, ***Black + NULL_LEAF = DB***; which symbolizes the deletion of a black color node that turns into a *NULL_LEAF*, then its replacement with another black node, which finally results in the formation of a DB node. The SA method, as demonstrated in [4] were symbolic arithmetic operations performed on **DB** nodes, their parent *p,* and sibling *s*. However, that research work did not cover the *rotation* of nodes when removing **DB**. This paper extends that work in [4] through a further refinement of the SA method for **DB** removal to make it adaptable to node rotations and tree balancing.

### 1.2 Node Color-Change and Tree Rotation

The deletion of a black node that forms a **DB** or its replacement node is where lies the complexity of ***rotation, recoloring*** and ***rebalancing*** (*3R*) of a RB tree. This paper demonstrates a new and easy approach for handling the complexity of the *3R* in RB trees. Our step-by-step procedure is the SA algorithm which involves the use of the symbols: **R** (for red), **B** (for **black**), and **DB** (for **double-black**) colors, respectively. Symbolically, **R**, **B,** and **DB** are color **operands**. In between the color operands are two operators, namely, "+" (**addition**), and "-" (**subtraction**). Depending on the case, to effect a color-change; the "+" and "-" operators are applied on the color operands in the process of balancing the black heights in a RB tree. Rotation is the first operation that is performed when a **DB** is formed in order to to balance a RB tree. From the theoretical (conventional) algorithm (TA) [5], rotations in a RB tree are determined by the case at hand, namely, LR, RL, LL, or RR rotation [1, 2, 6]. However, other factors to consider in the process of balancing a RB tree after a delete operation and the subsequent formation of a **DB** is whether the **DB** node has: *i) an inner, outer or both inner and outer black nephews;* or whether it has *ii) an inner, outer or both inner and outer red nephews*. The nephew of a DB is the child of the sibling of the **DB**.

### 1.3 Naming the Derived Symbolic-Algebraic Rule

The work of [4] demonstrates the application of the SA method in the removal of the **DB** node but did not cover the rotation of nodes. That work introduced, the subtraction of *-B* from **DB** as *-B(DB)*; the subtraction of *-B* from sibling *s* of **DB** as *-B(s)*, and the addition of *+B* to parent *p* of DB, *+B(p)* (Figure 2). The removal of DB with rotation is a complex operation. For DB nodes that involves rotation to be effectively deleted and to have a balanced tree; this paper extends our initial work of [4] by refining the SA methods to obtain three equation, namely

a) *General* Symbolic Algebraic Rule (***GSAR***),
b) *Partial* Symbolic Algebraic *Rule1* (***PSAR1***), and
c) *Partial* Symbolic Algebraic *Rule*2 (***PSAR2***).

From the point of a DB formation to the point where the tree is balanced; the given symbolic rules are deterministic, precise, and unambiguous such that we can list the stages and operations as they occur step-by-step. For instance, *i) if a DB is formed, ii) nodes are rotated to form a new sub-tree structure, iii) DB is removed using the general rule, iv) node colors are changed,* and *v) tree is balanced*. The concise operations contained in the SA algorithms can help students and teachers to quickly answer the **"***what***,** *how,* and *why***"** questions about DB removal in RB trees which were hitherto difficult.

### 1.4 Contributions

The formation of a **DB** node from the deletion of a black node or when a black leafnode replaces a deleted *internal* ***Red*** node causes a reduction in black heights --- the number of black nodes on any simple path in a RB tree. The conventional (traditional) algorithm **(TA)** for fixing the **DB** node problem is very complex and fluid and also challenging for a quick grasp in teaching and learning. This in turn affects students' understanding of **DB** removal. The contributions of this paper are therefore, 1) to create an unambiguous and concise process for removing **DB** nodes; 2) to propose a new method called the symbolic-arithmetic algebraic rules for the removal of **DB** nodes and balancing of RB trees; 3) to present three variants of the symbolic arithmetic rules for removing **DB** nodes and show their effectiveness on node color-change and tree balancing; 4) to establish a clearer procedure for **DB** node removal and RB tree balancing using mathematical process; and 5) to improve the teaching and learning of **DB** node removal in computer science curriculum

### 1.5 Statement of the Problem

The deletion of a black node or replacement of a black node by another black node unsettles the RB tree. The problems that are occasioned by a delete operation may include: i) a reduction in black heights, ii) tree rotation, iii) tree restructuring, and iv) recoloring of nodes. The traditional algorithm (TA) and research approaches geared towards simplifying the learning of **DB** removal pose a lot of doubt to students on, i) when to rotate a DB, ii) when to recolor a node, and iii) which node to recolor to rebalance black heights. Thus, our research work revolves around the statement that: *There is a mathematically-based system of approach that can be used to precisely and unambiguously analyze, state, and ease the learning of the traditional algorithm of double-black removal, recoloring and rebalancing of nodes in RB trees.*

This paper continues with related works in section 2. In section 3 we present our methodology and foundational theory of the concept of SA operation as well as the three derived SA algorithmic methods. Section 4 discusses the SA operation and **DB** removal using different RB tree structural examples. The section also presented our findings, graphical visualization of the findings, and students' feedback on their user-experience *w.r.t.* our SA method compared to the traditional algorithm (TA) [5] for **DB** removal. Section 5 is conclusions and further works.

## 2. Related Works

A RB tree is one of several binary search trees. In data structures, every binary tree has their peculiar properties that define, namely; its structure, height and relationships between nodes, and the tree's stable state. One of the attributes of the RB tree is the "*color*" field through which every node in a RB tree is assigned a color which is either red or black [1, 2, 7].

### 2.1 Properties of Red Black Trees

   a)   Each node is either red or black.
   b)   The root node is black.
   c)   Each leaf node (i.e. NULL_LEAF) is black.
   d)   The children of a red node are black.
   e)   For a node, all simple paths from the node to descendent leave contain the same number of black nodes.
   f)   Two consecutive nodes cannot be both red.
   g)   A red black tree is a binary search tree (BST).

At the point where all these requirements are satisfied, a RB tree is created. The aforementioned properties defined the stable or balanced state of the RB tree. Furthermore, if the insert or delete operation is called on a given node, the structure of the tree and the nodes' color also change to reflect the requirements which helps the tree to rebalance itself [8]. Based on node coloration; the children of a black node can; *i) both be blacks, ii) red and black, or iii) both reds* (Figure 1). But for a red node, its two children must always be black [9]. This is because a red node must have a black parent node as well as a black child node. Otherwise, the property that states *"no two nodes that are connected side-by-side with an edge can be red"* would be violated.

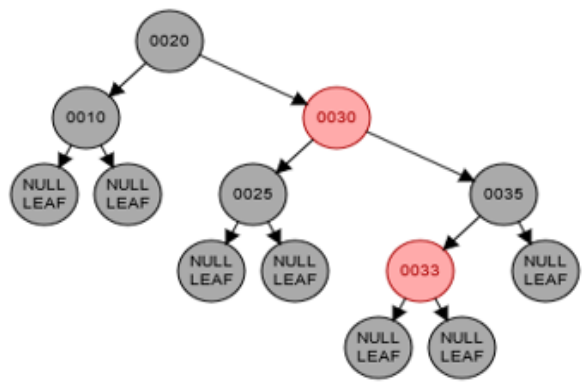

Fig. 1. A balanced red-black tree showing null leaves. In a Red-Black Tree Visualizer [10].

After operations like insert, update or delete; tree-based data structures such as the AVL and RB trees must be balanced [11]. A RB tree is balanced if a simple path from the root node to every descendant leafnode has equal number of black nodes (Figure 1). On the other hand; it is unbalanced if there are unequal number of black nodes on every simple path from the root to the leafnodes. The RB tree is a famous data structure for storage and management of data. In the non-teaching fields, several studies e.g. [12, 13, 14, 15] have been conducted on the impact analysis of RB trees on memory management and performance. For instance, [16] conducted a research on the application of RB trees in wireless sensor networks. The use of RB trees for optimising costs in network trees with a prescribed algorithm [17]. As an abstract data type that is aimed at understanding computer storage and software engineering; the operations performed on a RB are insertion, rotation, deletion, and recoloring of nodes.

The RB tree is an important topic in data structures in the study of computer science. Students can find RB trees very challenging and difficult to learn [9, 18, 19]. In the lookout for new approaches to teaching RB trees, Wu et al. [19] proposed a customized platform for learning data structures. After the proposal of a series of steps to learn data structures with-ease there was however no mention of tangible topical areas in the subject of data structures. To improve the efficiency of teaching data structures, [20] conducted some experiments using different student groups e.g. Group1 vs. Group2 to compare the effectiveness of his teaching strategy using *visualization tool vs. flipped-classroom* in the subject area of Hashing and Trees (BST, RBT, AVL). The scores obtained in the research were used to determine what group learned the most. King [21] reported how their university data structures & algorithms curriculum was redesigned to reflect experimental analysis as against only theory of concepts in data structures. The revised curriculum and practice, thereafter as reported, enabled students to include experimental analysis in their studies to connect computer science theory with software engineering practice. The methodology in the report lacked algorithmic technique, and the report concluded its findings using the Likert Rating Scale for the collection of metrics.

The report of [22] presented a list of the most conventional topics in data structures & algorithms and emphasized the importance of logic in algorithmic studies --- that algorithms are *logics*. Furthermore, that data structures curriculum needs to be supported with critical computational thinking and *formal proofs* and *logic*. In [9] a top-down insertion method was proposed to address the problem of single and double rotation through a granularity approach to balance a RB after an insert operation. The granularity approach prescribed a step-by-step selection of rules for students to follow and balance a RB tree. But the aforementioned approaches for addressing the teaching and learning of RB trees that involves rotation and recoloring of nodes e.g. [9, 23] has not made the problem any less or easier.

The insert operation of a new node in a RB tree is the simpler of two hard problems. The harder is dealing with the deletion of a black node or the replacement node [24] and the subsequent rebalancing of the tree. In order to address the removal of **DB** node, [6] used formalized statements in the description of nodes' color, recoloring and tree balancing which (in any way) did not simplify the learning challenges faced by students in the learning of **DB** removal algorithm. In [24] a parity-seeking delete algorithm was introduced with a goal that is similar to the aim of our work in this paper. They introduced a pedagogically sound and easy way to understand the algorithm for RB tree deletion algorithm. The rationale which was to balance a RB tree by repairing a defective subtree also left a bit of complexities in learning and understanding. Also, the work of [25] presented 2-3 variants RB tree, called the Left-Leaning Red-Black trees (LLRB) and proposed a concise number of deletion algorithms. However, the deletion algorithm of the LLRB was still complex and was not suitable for educational use.

Based on these complexities that are left with the deletion algorithm of the RB tree, the aim of this paper is to simplify the learning, understanding, and teaching of **DB** node removal using basic symbolic-arithmetic operations after a delete operation. Students have expressed satisfaction in the use of the SA algorithmic rules over the TA algorithm in that the TA is not deterministic nor is it precise in approach for ease of understanding. See the relevant Chapters in [1, 2] and RB tree configuration in visualization tools [26]. Hence, the quest for this new SA approaches.

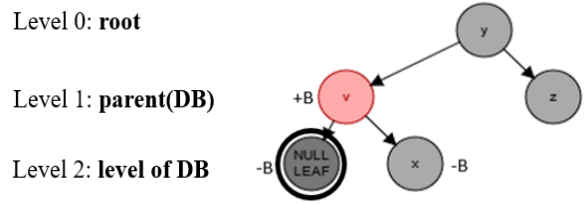

Fig. 2. The symbolic ***subtraction*** of the *color black* (represented as *-B*) from two siblings and the ***addition*** of the color *+B* to their *parent v*. Thus, -B(leftChild) ×-B(rightChild) = +B(parent). [4]

### 2.2 The Symbolic-Algebraic Arithmetic Rules

In [4], the SA operations are stated as

$$black + black = \textbf{\textit{double black}} \tag{1}$$

$$\textbf{\textit{double black}} \text{ - } black = black \tag{2}$$

$$black \text{ - } black = \textcolor{red}{red} \tag{3}$$

$$\textcolor{red}{red} + black = black \tag{4}$$

$$\textcolor{red}{red} + \textbf{\textit{double black}} = black \tag{5}$$

$$\textcolor{red}{red} \text{ - } black = black \tag{6}$$

$$\textbf{\textit{null(DB)}} - black = NULL\_LEAF \tag{7}$$

The given symbolic operations imply the simple **addition** or **subtraction** of *red* or *black* color *to* or *from* an existing node color to obtain the resultant color in the process of **DB** removal and rebalancing of black heights. *Note that the color operands in the formulas above cannot be moved to the other side of the equality sign, otherwise, conflicts may occur.*

### 2.3 The Change Factor

The change factor is the color that is added to or subtracted from an original color. For example, if a node is initially black and an extra black color is added to it, the from Equation (1), we have

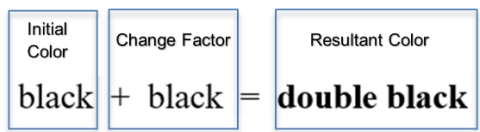

Fig. 3. An example of the change factor in Equation (1). [4]

### 2.3.1 The Transmission Rules of The Change Factor

a) The path, or order, of the recoloring process travels from the deleted node to its adjacent node, both along and against the arrow is allowed (*see* connected nodes, Figure 2).

b) The transmission (of the SA rule operation) stops when a tree is balanced; but continues if another **DB** node is regenerated as a result of the SA operation.

## 3. Methodology

This paper presents three refined symbolic arithmetic (SA) algorithm based on the previous works of [4] for **DB** node removal. In this presentation, we have adopted the use of formal logic in the form of $F : X \longrightarrow Y$ to depict the operations on the nodes in a RB tree that involved rotation of nodes when removing a **DB** node. The use of formal logic for information representation has enriched the understanding and learning of the **DB** removal problem, and it has made the process clearer, precise, concise and explainable. Logic based explanations are still present in modern day systems. As stated in [21], formal logic should be used to support the learning and understanding of data structures, because logical formula supports explainable systems [27]. Our mathematical approach is well simplified and to a very large extent it is also different from other approaches like the use of logical formulas in [28] or functional programming in [29]. Just as formal logic has found its place in modern explainable artificial intelligence (XAI) [27] systems so it continues to support software engineering systems as exemplified in this work for teaching the **DB** node removal.

The three SA rules that we have developed for **DB** removal revolves around the operation of node recoloration, and tree balancing and they are namely

1) **General** *symbolic arithmetic* **rule** *(GSAR)*,
2) **Partial** *symbolic arithmetic* **rule1** *(PSAR1)*,
3) **Partial** *symbolic arithmetic* **rule2** *(PSAR2)*.

In RB trees, there are four different types of rotation which are LR, RL, LL, and RR rotation. From our observation in this work, the rotation of nodes in the process of balancing a RB tree is determined by the ***position*** and ***color*** of the following nodes

- the *nephew of the DB*,
- *parent of the DB*,
- *sibling of the DB*, and
- the *grandparent of the nephew* (which is also the *parent of the DB*)

Consequently, we are using the following parameters to denote the aforementioned nodes such that

- ***u*** is the *deleted* node (or value) that formed the ***DB*** i.e. ***deleted*(*u*) ⟹ *DB*(u)**
- ***DB*** is the node that must be removed
- ***p*** is the *parent* of DB, ***p(DB)***
- ***s*** is the *sibling* of DB, ***s(DB)***
- ***r*** is a **special** *nephew* of DB, ***r(DB)* :** and we called this the **VIP** (very important personality) node
- ***x*** is a second *nephew* of DB, ***x(DB)*** : Thus, ***r*** and ***x*** are cousins; and are children of sibling, ***s***, of the **DB**
- ***g*** is the *grandparent* of ***r***, ***g(r);*** and of ***x***, ***g(x)*** respectively. Thus ***g(r) ≡ p(p(r))*** and ***g(x) ≡ p(p(x))***

Secondly, by correctly identifying the type of rotation i.e. LL, RR, LR and RL as well as the rotation of nodes in their proper order, say ***from LR to LL*** or ***from RL to RR***, we observed that any two of the three SA rules can be combined to effectively resolve the *3R* problems in RB trees. The combination order of any two SA rules is the pairs of

a) **GSAR & PSAR1**.
b) **PSAR1 & GSAR**.
c) **PSAR1 & PSAR2**.
d) **GSAR & PSAR2**.
e) Or just the **GSAR** originally from the work of [4].

### 3.1 Theoretical Foundation

Respectively, Figures 4 and 5 are basic RB tree problems with **DB** formation. To resolve the 3R problem, we have used a step-by-step transition method to process the changes that occurred in different stages. In the aforementioned Figures, the nodes have been theoretically given the values of v, x, y, and z, respectively. The RB tree is a *Binary Search Tree* (BST). Thus, these labels are guided by the property of the BST. A RB tree that has a DB node with any of the four types of structure of rotation i.e. LR, RL, LL, or RR problem must, firstly, be rotated before the application of the SA rules. The step-by-step stages, namely, a), b), c), and d) depicts the procedure that we have established in applying the SA rules.

For instance, in Figure 4, the pair of rules to establish the 3R algorithmic procedure is the ***GSAR & PSAR1***. In Figure 4(a), the **DB** node has a black parent, ***B(p(DB(v)))***, and ***two black nephews***, namely, inner ***x, B(innerNephew(x)),*** and outer ***z, B(outerNephew(z))***. While ***x*** is the *black inner nephew* to DB that can be stated as ***B(innerNephew(x(DB)))***, ***z*** is then given as the ***B(outerNephew(z(DB)))***. With two nephews to a DB, one of the nephews must be a **VIP** node. In this color configuration that involves five connected and concerned nodes to a **DB** i.e. the **DB**, ***B(p(DB))***, ***R(s(DB))***, ***B(innerNephew(x(DB)))***, and ***B(outerNephew(z(DB)))***; our interest will be on the ***B(innerNephew(x(DB)))***. Thus, where there are **two black nephews** to **DB**, we will be interested in changing the color of this ***B(innerNephew(x(DB)))***. The ***B(innerNephew(x(DB)))*** is the node that we have called the **VIP** (very important personality) node. It is the **VIP** node because it is ***not*** the node that determines the case (of rotation), in this case a **RR**; but it is the node whose color would change from **B(x)** to **R(x)** at stage 4(c) of Figure 4.

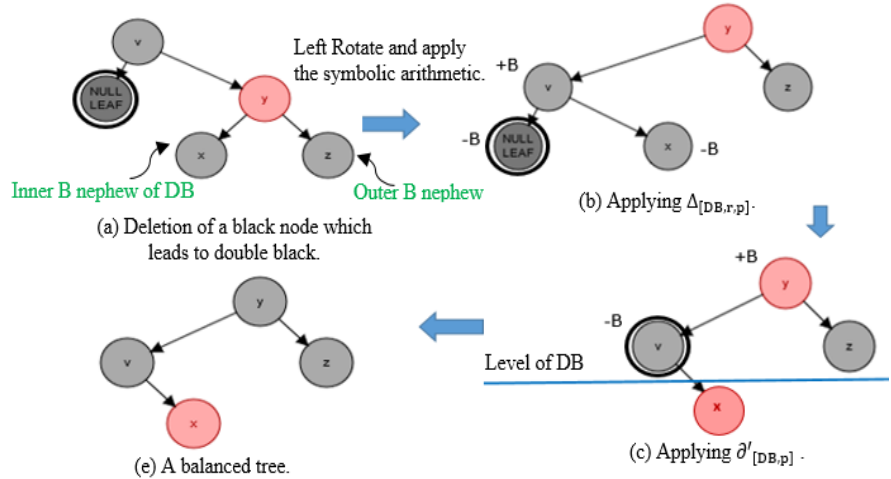

Fig. 4. (a) Depicts a RR case with a DB from a deleted black node. (b) Shows a left rotation on *node v* --- the parent of DB, and the operation of the *general rule* $\Delta_{[DB,r,p]}$; (c) The depicts the *partial rule1* $\partial'_{[DB,p]}$. (d) A balanced tree.

In Figure 4(a), based on the case which is a RR, we left-rotate the ***B(p(DB))***. Then, in Figure 4(b), we apply the *general rule* **GSAR**, $\Delta_{[DB,r,p]}$. The GSAR rule depicts the the ***subtraction*** of the *color black, -B,* from both the **DB** and its *new sibling x* which was initially its *black inner nephew*; and then the ***addition*** of the color black, +B, to their ***parent v, B(p(v))***. Symbolically, this process is depicted as, ***-B(DB), -B(innerNephew(x(DB)))***, and ***+B(p(DB))***. Now, in Figure 4(c) we have a ***new DB*** as a result of the addition of black, +B, to the ***B(p(DB))***. At the new level of **DB** which is one level up the RB tree, we then apply the second rule, that is the *partial rule1* **PSAR1**, $\partial'_{[DB,p]}$, by subtracting one black, *-B*, from the **new DB(v)**, ***-B(newDB(v))***; and adding one black, +B, to the **Red** *parent y* of the ***new DB***, ***+B(R(p(newDB(y))))***, while the new *sibling z* of the **new DB** is **exempted**. In Figure 4(b) and Figure 4(c) the **GSAR** and the **PSAR1** are symbolically denoted as $\Delta_{[DBR,r,p]}$ and $\partial'_{[DB,p]}$, respectively. Based on the foregoing analysis, we arrived at ***Corollary 1***.

▪ ***Corollary 1*:**

*In a RB tree, if there are **two black nephews** to a DB with a black parent, then the case is a **LL** or **RR** case decided by the **black outer nephew** but the **VIP** node would be the **black inner nephew** that will have a color change.*

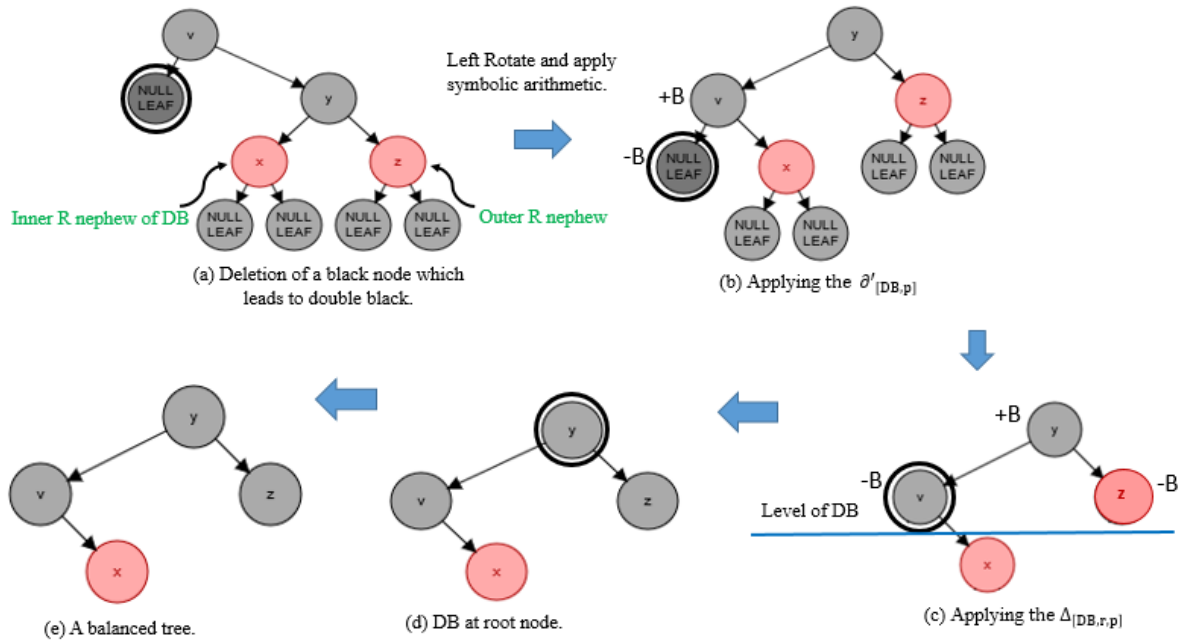

Fig. 5. (a) Shows a RR case with a DB from a deleted black node. (b) Depicts a left rotation on *node v* and the operation of *partial rule1* $\partial'_{[DB,p]}$. (c) The *general rule* $\Delta_{[DB,r,p]}$. (d) A balanced tree given that ***node y*** is a root node.

In Figure 5(a), the **DB** also have a black parent $p$ that is given as $B(p(DB(v)))$ but *two red nephews*, $x$ and $z$. While $x$ is the **Red inner nephew** given as $R(innerNephew(x(DB)))$; $z$ is the **Red outer nephew** that is given as $R(outerNephew(z(DB)))$. In this type of color configuration i.e. a black parent of **DB**, $B(p(DB))$, and two red nephews, $R(innerNephew(x(DB)))$ and $R(outerNephew(z(DB)))$; our interest will be on the $R(outerNephew(z(DB)))$ to completely resolve the 3R problem. In Figure 5(b) the first rule to apply will be the *partial rule1 PSAR1*, $\partial'_{[DB,p]}$. As mentioned earlier, with two nephews to a **DB**, one of the nephews must be the **VIP** node. In this case the VIP node is the **outer red nephew** $z$ to $DB$, $R(outerNephew(z(DB)))$. As such, we will **subtract** one black, **-B**, from the **DB**, **-B(DB)**; and **add** one black, **+B**, to its **parent v**, **+B(p(DB))**. Note that the **red inner nephew** $x$ to **DB** that is given as $R(innerNephew(x(DB)))$ is **exempted** from this arithmetic operation. In Figure 5(c) we have a **new DB** which is one level-up the tree. At the new level of the **DB**; we will apply the *general rule* by subtracting one black, **-B**, from both the **new DB(v)** and its sibling which was the **original** *red outer nephew* $z$ to *DB*, respectively, and then **add** one black, **+B**, to their new parent $y$, **B(y(DB))**. Symbolically, this is stated as **-B(newDB)**, **-B(R(outernephew(z(DB))))**, and **+B(p(y(DB)))**. The pair of rules applied in this algorithmic procedure are the **PSAR1 & GSAR**. In Figure 5(b) and Figure 5(c), the **PSAR1** and **GSAR** are shown as $\partial'_{[DB,p]}$ and $\Delta_{[DB,r,p]}$, respectively. Based on the foregoing analysis, **Corollary 2 was drawn**.

▪ **Corollary 2**:

*In a RB tree, if there are two red nephews to a DB with a black parent, then the case is a **LL or RR case** decided by the red outer nephew but the VIP node would be the red outer nephew that will have a color change.*

Note that the **VIP** nodes such as the **B(innerNephew(x(DB)))**, (in Figure 4); and the **R(outerNephew(z(DB)))** to *DB* (in Figure 5) are VIP-nephew nodes of the two nephews to **DB** to have a color change. In this paper, the **VIP** nodes are hereafter labeled as the **node r** in all the examples and illustrations. In the *general rule* and *partial rule2*, respectively, they are represented by the parameter $r$. At every stage of the SA operation, the algorithm tests whether the RB tree is balanced or not. If the RB tree is balanced, the process of further rotation or SA application ends. Otherwise, we recur one step, further up, the tree as shown in Figures 4(c) and 5(c). In the following section, we present the detailed procedure of the three SA rules. We have used the RB tree **Case 2** and **Case 3** configuration to illustrate these procedures. The complete expression of the **general rule** and the *two partial operations* as denoted in the **partial 1** and **partial 2** rules, is given as

$$general\ rule = partial\ rule1 + partial\ rule2$$

and mathematically as

$$\Delta_{[DB,r,p]} = \partial'_{[DB,p]} + \partial''_{[r]} \tag{8}$$

## 3.2 General Rule: GSAR

The *general rule*, *GSAR*, denoted that is denoted as $\Delta_{[DB,r,p]}$ is the rule that applies to a **DB** node and its sibling $s$, and to their parent $p$. Thus this rule causes a color change to a parent node and its two children. The **GSAR** operates on three inter-connected nodes as shown by the **delta** $\Delta$ symbol that has 3 vertices. The vertices indicate all the three nodes that are involved in the arithmetic operation. Since we are dealing with a **DB** problem, one of the three connected nodes must be a **DB**, and the **DB** must be one of the two children of a parent $p$. In logical form, this rule is given as

$$\forall_{[DB,r,p]} : \Delta_{[DB,r,p]} \rightarrow -B_{DB} \wedge -B_r \wedge +B_p \tag{7}$$

That is

$$\forall_{[DB,r,p]} : \Delta_{[DB,r,p]} \rightarrow (DB - B) \wedge (r(DB) - B) \wedge (p(DB) + B) \tag{8}$$

which states that, for all **DB**, **r**, and **p**, to remove the DB *w.r.t.* **p(DB)**, **r(DB)**; subtract **-B** from the **DB**, and its *nephew **r***; and add **+B** to parent **p** of **DB**. This translates to the symbolic arithmetic form

```
DB − B ⟹ NULL_LEAF
B(r) − B ⟹ R(r) || R(r) − B ⟹ B(r)
B(p) + B ⟹ newDB(p) || R(p) + B ⟹ B(p)
```

Particularly, where the operation $DB - B = NULL\_LEAF$; the **DB** must have been formed from a deleted black node. Given the *property(object)* formalism; the notation $B(r)$ states that the nephew $r$ has the color black $B$ (or that $r$ is black), and $B(p)$ means that the parent $p$ is black. It is important to state that the **resultant color** of a node after a subtract, $-B$, or an add, $+B$, operation is dependent on the existing color of that node. For instance, the resultant color $B(p)$, from the operation $R(p) + B = B(P)$. The pre-existing color on the node $p$ is *Red*, $R$. This is further illustrated with conditional *IFs* as

```
IF R(p) THEN
    B(p) = R(p) + B
IF B(p) THEN
    newDB(p) = B(p) + B
```

### 3.2.1 Characteristics of The General Rule

i.   The *general rule* is operated on three connected nodes.
ii.  The *general rule* uses a *-B* symbolic operator on two children and a *+B* on their parent.
iii. The *general rule* applies to a **DB** whose parent can be a *Black* node, $B(p(DB))$; or *Red* node, $R(p(DB))$, at the time of formation of the **DB**;
iv.  The *general rule* works with either the *partial rule1* or the *partial rule2* to balance a RB tree.

## 3.3 Partial Rule1: PSAR1

The *partial rule1*, denoted as $\partial'_{[DB,p]}$ is the rule that partially applies to two nodes at any given instance. From the notation $\partial'_{[DB,p]}$, this rule **exempts** one node of the three nodes that can be handled by the *general rule, GSAR*. Hence the name *partial* from partial differential calculus**.** The *partial 1* is the first of the two extended partial rules that we are projecting in this work. The *partial rule1* causes color-change on a **DB** and the parent $p$ of **DB** as given in the notation $\partial'_{[DB,p]}$. In this operation, the **DB** may be the *original* **DB** and $p$ the parent of the **DB** at the point of formation of the **DB**; or a *new* **DB** and its new parent $p$ after the *general rule*, $\Delta_{[DB,r,p]}$, operation. The single prime ,'; symbol in the notation $\partial'_{[DB,p]}$ indicates the node that is exempted in the operation of this rule. In logical formalism , this is given as

$$\forall_{[DB,p]} : \partial'_{[DB,p]} \rightarrow -B_{DB} \wedge +B_p \tag{9}$$

That is

$$\forall_{[DB,p]} : \partial'_{[DB,p]} \rightarrow (DB-B) \wedge (p(DB)+B) \tag{10}$$

This states that, for all **DB** and its parent $p$; we can remove the **DB**, by calling the *partial rule1* function which subtracts *-B* from the **DB**; and add *+B* to the parent $p$ of the **DB**. Now, the color of the *p(DB)* could be black $B$ or red $R$. In symbolic arithmetic (SA) form, we have

```
B(p) + B ⟹ DB(p) || R(p) + B ⟹ B(p).
```

The operation of the *partial rule1*, $\partial'_{[DB,p]}$, is dependent on two factors, which are namely, *i) the position of the inner and outer nephews, r and x*; and *ii) the color, whether Red or **Black**, of the nephews r and x* of the DB, respectively (*see Figure 4 and 5*).

### 3.3.1 Characteristics of The Partial Rule1

i.   The *partial rule1* operates on a **DB** that may have a black parent or a red parent: $B(p(DB))$ or $R(p(DB))$.
ii.  The *partial rule1* can be performed before the *general rule* or vice versa: This is dependent on the *p(DB)* and the **color** and **position** of the (*VIP node*) *nephew r* of **DB**, $r(DB)$.
iii. *Partial rule1* denoted as $\partial'_{[DB,p]}$ always comes first before the *partial rule2* that is denoted as $\partial''_{[r]}$ (*next section*).
iv.  *Partial rule1* keeps one node color constant while it makes changes to the other second nodes' color, *see Figure 4 and 5*.

## 3.4 Partial Rule2: PSAR2

The *partial rule2* which is denoted as $\partial''_{[r]}$ is an operation that is performed on only one node in an unbalanced RB tree. From the notation $\partial''_{[r]}$, this rule **exempts** two nodes of the three connected nodes that are depicted by the *general rule* $\Delta_{[DB,r,p]}$. The double prime ,''; symbol in the notation $\partial''_{[r]}$ indicates the two exempted nodes. This rule applies to a **DB** node that has a *Red* parent $p$, i.e. $R(p(DB))$. In logic form, we state that

$$\forall_{[r]}: \partial''^{[r]} \rightarrow -B_r \tag{13}$$

That is

$$\forall_{[r]}: \partial''_{[r]} \rightarrow (B(r) - B)) \,||\, (R(r) - B) \tag{14}$$

That, for all *r* that are the **original nephews** of **DB**; to remove a **DB** and balance the tree, we then subtract **-B** from the *original nephew r* of **DB**. The node *r* is the *nephew* of a **DB** at the point of formation of the DB. Note that as the tree undergoes transformation as a result of node rotation and the application of the symbolic arithmetic (SA) operation on the respective nodes; the position and color of the *original nephew r* to the **DB** also changes, *see Figures 13, 14 and 15*. At this point, the *partial rule2*, $\partial''_{[r]}$, will be acting to change the *new and current color* of nephew *r*; *not the original* color of *r*. In symbolically arithmetic

$$\mathbf{B}(r) - \mathbf{B} \implies R(r) \,||\, R(r) - \mathbf{B} \implies \mathbf{B}(r)$$

In conditional statements, we state that

**IF B**(r) **THEN**
　　R(r) = **B**(r) − **B**
**ELSE**
　　**B**(r) = R(r) − **B**

### 3.4.1  Characteristics of Partial Rule2

i.　The *Partial rule2* is performed on only one node.
ii.　The *Partial rule2* uses only one symbolic operator which is the subtraction operator, **-B**, from a node.
iii.　The *partial rule2* is always a secondary rule to the *general rule* or *partial rule1*, respectively.
iv.　The *Partial rule2* comes into effect in a tree structure where a **DB** has a **Red parent p, R(p(DB))**, at the point of formation of the **DB**.

### 3.5  Procedure of the Symbolic Arithmetic Rules

The following section presents the step-by-step procedure for removing a **DB** node using the SA algorithmic rules. *Procedures 1, 2,* and *3* describe the process for a DB that has a black parent *p, B(p(DB));* and *Procedures 4* and 5 for a **DB** node that has a **Red** parent *p, R(p(DB)).*

**Procedure 1 [Figure 11]:** *LL case* and *RR case:* **Black Parent** *of DB* and **Two Black Nephews** *of DB*
　**1. Start**
　2. **IF** a *black node* is deleted and the deleted node has a **black parent** and **two black nephews**, *r & x*:
　**3.　WHILE** tree is not balanced:
　4.　　　Rotate the parent, p, of DB
　5.　　　Apply the *general rule* $\Delta_{[\mathbf{DB,r,p}]}$ :
　6.　　　　$\forall DB, r, p : \Delta_{[DB,r,p]} \rightarrow -B_{DB}\,\Lambda - B_r\,\Lambda + B_P$
　**7.　　　END** *general rule*
　8.　　　Apply *partial rule1*, $\partial'_{[\mathbf{DB,p}]}$, at the level of DB:
　9.　　　　$\forall DB, p : \partial'_{[DB,p]} \rightarrow -B_{DB}\,\Lambda + B_P$
　**10.　　END** *partial rule1*
　11.　　　IF the new DB is root:
　12.　　　　root $\implies B$　/** root is always black */
　13.　　　Tree is balanced
　**14.　END WHILE**
　15. **END**

**Procedure 2 [Figure 8 & 9]:** *LL Case* and *RR Case:* **Black Parent** *of DB* and *Two Red Nephews of DB*
　**1. Start**
　2. **IF** a *black node* is deleted and the deleted node has a **black parent** and *two red nephews*, *r & x*:
　3.　**WHILE** tree is *not balanced*:
　**4.　　　Rotate** the *parent*, *p*, of DB
　5.　　　Apply *partial rule1*, $\partial'_{[\mathbf{DB,p}]}$, at the level of DB:
　6.　　　　$\forall DB, p : \partial'_{[DB,p]} \rightarrow -B_{DB}\,\Lambda + B_P$
　**7.　　　END** *partial rule1*
　8.　　　Apply *general rule* $\Delta_{[\mathbf{DB,r,p}]}$ :

9.        $\forall DB, r, p : \Delta_{[DB,r,p]} \rightarrow -B_{DB}\,\Lambda - B_r\,\Lambda + B_P$

**10.**    *END general rule*

11.    **IF** the new DB is root:

12.       root $\Longrightarrow B$  /** root is always black */

13.    Tree is balanced

**14.**  **END WHILE**

15. **END**

---

***Procedure 3 [Figure 6 & 7]:*** *LR Case and RL Case:* **Black Parent** *of DB and* *One Red Inner Nephew of DB*

  **1. Start**

  2. **IF** a *black node* is deleted and the deleted node has a ***black parent*** and *one red inner nephew, r*:

  3.  **WHILE** tree is not balanced:

  4.    **Rotate** the *sibling, s,* of DB

  5.    Apply the *general rule* $\Delta_{[\mathbf{DB,r,p}]}$ at the level of DB:

  6.      $\forall DB, r, p : \Delta_{[DB,r,p]} \rightarrow -B_{DB}\,\Lambda - B_r\,\Lambda + B_P$

  **7.**    *END general rule*

  8.    **Rotate** the *original* parent, p, of DB  /** new DB */

  9.    $DB_{new} - B \Longrightarrow B$  /** root is always black */

  10.   Tree is balanced

  **11.**  **END WHILE**

  **12. END**

---

***Procedure 4 [Figure 13 & 14]:*** *LR Case* and *RL Case: Red Parent of DB and One Red Inner Nephew of DB*

  **1. Start**

  2. **IF** a *black node* is deleted and the deleted node has a *red parent* and *one red inner nephew, r*:

  3.  **WHILE** tree is not balanced:

  4.    **Rotate** the *sibling, s,* of DB

  5.    Apply the *general rule* $\Delta_{[\mathbf{DB,r,p}]}$ at the level of DB:

  6.      $\forall DB, r, p : \Delta_{[DB,r,p]} \rightarrow -B_{DB}\,\Lambda - B_r\,\Lambda + B_P$

  **7.**    *END general rule*

  8.    **Rotate** the original parent, p, of DB

  9.    Apply *partial rule2,* $\partial''_{[r]}$, at the level of DB:

  10.    $\forall r : \partial''_{[r]} \rightarrow -B_r$

  **11.**    *END partial rule2*

  **12.**    Tree is balanced

  **13.**  **END WHILE**

  **14. END**

---

***Procedure 5 [Figure 15]:*** *LL Case* and *RR Case: Red parent of DB and One Red Outer Nephew of DB*

  **1. Start**

  2. **IF** a *black node* is deleted and the deleted node has a *red parent* and *one* red *outer nephew, r*:

  3.  **WHILE** tree is not balanced:

  4.    **Rotate** the *parent, p,* of DB

  5.    Apply the *partial rule1* $\partial'_{[\mathbf{DB,p}]}$ at the level of DB:

  6.      $\forall DB, p : \partial'_{[DB,p]} \rightarrow -B_{DB}\,\Lambda + B_P$

  **7.**    *END partial rule1*

  8.    Apply *partial rule2,* $\partial''_{[r]}$:

  9.    $\forall r : \partial''_{[r]} \rightarrow -B_r$

  10.   *END partial rule2*

  11.   Tree is balanced

  **12. END WHILE**

  **13. END**

## 4. Discussions

    In this section we present our technical but simple SA methods for the removal of **DB** nodes with rotation. The illustrative examples demonstrate the *general rule, partial rule1* and *partial rule2* for the LR, RL, LL, and RR cases of the RB tree. In a BST and by extension RB trees, a *parent* is a node that has two children: the ***leftChild*** and ***rightChild***. Based on the examples provided in this work, the first step to identifying any of the four cases of rotation is to locate the parent *p* of the ***DB***, *p(DB)*. Thereafter, start putting labels to the nodes from the *p(DB)* down through the edges to the

sibling *s* of **DB**, *s(DB)*; and through to the nephews *r* and *x* of the DB i.e. *r(DB)* and *x(DB)*. In Figure 6 for example, we knew that Figure 6(a) is a **LR** Case by navigating from **node 40**, which is *p(DB)*; down to the black **node 20**, which is the *B(s(DB))*; and finally to **node 30**, the red inner nephew *r* of DB, *R(r(DB))*. The node *B(s(DB))* is a *leftChild* of the *p(DB)* and *R(r(DB))* is the *rightChild* of the *B(s(DB))*. Hence, the LR Case structural navigation.

### 4.1 Recursive Subtraction and Addition of Colors

Irrespective of whether a DB node is a *leftChild* or *rightChild* of its parent, our SA algorithm *subtracts 1 black color -B* from both children of the same parent *p* where one of the children is a **DB** and *adds 1 black color +B* to their parent. Such as *-B(leftChild)*, *-B(rightChild)* and *+B(p(DB))*. Technically, the strategy is a simple addition of *+B* color to/from an existing *Black B* node or to *Red R* node e.g. a parent node; or the subtraction of *-B* color from an existing *Black B* or *Red R* node e.g. from child nodes. Always, the SA algorithm always tests whether a RB tree is balanced or not in the process of addition or subtraction of color *B*. If the RB tree is not balanced, then we recur up the tree by calling on the next appropriate rule as discussed in the *methodology*. As we apply the SA rules and recur up the tree, we must apply the *GSAR* $\Delta_{[DB,r,p]}$ and the *PSAR1* rules at the level of the DB node, and the *PSAR2* $\partial''_{[r]}$ on only the nephew node, *r*. Also, as we recur up the tree, the traversal must be upward along the path of a given **DB** node towards the root of the tree. This process recursively continues if there is a re-occurrence of a **DB** by virtue of *+B* addition to another black *B* node which in turn becomes a new parent *p*. But if the new **DB** *p* is a root node then it turns a *single black B* because a root node is always black.

### 4.2 PART I: Examples of DB Removal with a Black Parent

For every RB tree case, the following examples demonstrate the detailed step-by-step process of balancing a RB tree without violating its properties. In this SA operation, we have chosen to use the parameters **u**, **DB**, **p**, **s**, **r**, **x**, and **g** as already described in our methodology. Also, in algorithmic representation; we shall use the following notations, such as, *B(p)* for a black parent node, *DB(u)* for the double-black node *u*, *B(s)* and *R(s)* for the black or red sibling of a **DB** node. For a much more detailed representation, wherever necessary, we would state *B(s(DB))* as the black sibling, *R(s(DB))* for the red sibling, and *B(p(DB))* for the black parent *p* of DB, respectively. Also, others are *R(innerNephew(r(DB)))* for the red inner nephew *r* and *B(outerNephew(x(DB)))* for the black outer nephew *x* of DB, respectively. Furthermore, as mentioned earlier, the parameter *r* represents a *special node* called the **VIP** *nephew* of a DB, *r(DB)*, *see section 3*. In addition, *g(r)* shall be used to represent the grandparent *g* of the nephew *r* of a DB, and *g(r)* is equivalent to *p(p(r))*, the parent of the parent of nephew *r*, and *B(g(r))* as the black grandparent of *r*. The parameters *p*, *u*, *s,* and *r* shall be assigned their values whenever necessary in the process of evaluating the SA algebraic algorithm. For instance, where a black parent node value is *40*, this will be written as *B(p(40))*. It is important to note that **DB** nodes are NOT only formed from the deletion of black leaf nodes but are also formed in the position of a black leaf node that has moved to replace a deleted internal red node.

#### 4.2.1 Left-Right Case of Double-Black

**Example 1**: *The Case of a DB u with a black parent and one red inner nephew r.*

Our first demo is Figure 6 which presents a **DB** node formation after the deletion of the node value that is labeled *u*. To resolve the 3R problem of **DB** removal, firstly, we have to identify the Case. Starting from the *DB(u)*, to its black parent, *B(p(40))*, then to the black sibling *s* of **DB**, *B(s(20))*, and down to the red inner nephew *r*, *R(innerNephew(r(30)))*; the case of a DB with only *one red inner nephew, r(30)*. This is the type of nephew we have called the **VIP** *nephew r*. It a **VIP** because it had its color changed from red, *r(30)* to black *r(30)*. In addition, the **VIP** nephew *r* also decided the case of rotation to be LR.

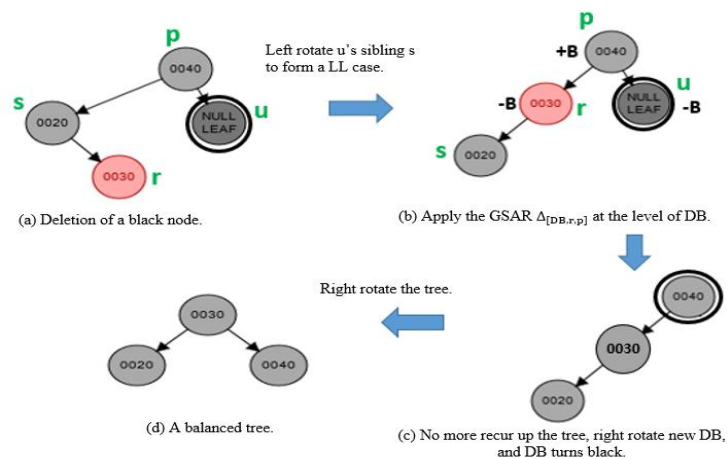

(a) Deletion of a black node.

(b) Apply the GSAR $\Delta_{[DB,r,p]}$ at the level of DB.

(d) A balanced tree.

(c) No more recur up the tree, right rotate new DB, and DB turns black.

Fig. 6. A LR case after the deletion of a black node.

We called the red *nephew*, *r*, the **VIP** node because it is the nephew node that changes its **color** at the **end** of the DB removal process or the **3R** (removal, recoloring, and rebalancing). More so, if a DB has two nephews, namely, *r* and *x*; amongst the two nephews, the **VIP** nephew *r* in this work is the only nephew that would change color. We shall see this in the forthcoming examples.

As a LR Case (Figure 6), we left-rotate the sibling *s* of DB, *B(s(20))* to have a LL Case. Then at the level of DB, we apply the *general rule*, $\Delta_{[DB,r,p]}$; to have a *new* **DB** at node *B(p(40))*. We then right-rotate the *new* **DB**, *DB(40)* to remove the DB. At this point, the tree is balanced i.e. *blackHeight = 2* (the equal number of black nodes along every simple path from the root to any descendants leafnode). Using algebraic notation, we describe this Case and the stages of the 3R resolution from the point of the **DB** formation to tree balancing as $DB_{LR}^{B(p),InnerR(r)}$. In a list format, we enumerate the operations and the different stages as

$$DB_{LR}^{B(p),InnerR(r)} = [DB, leftRotate(s(DB)), LL, \Delta_{[DB,r,p]}, rightRotate(newDB), remove\ DB] \quad (15)$$

and in first-order logic representation *w.r.t.* the red **VIP** nephew *r* and the black grandparent of *r*, *B( g(r))*, we state that

$$\forall g, r : B(g(r)) \wedge {\color{red}R(r)} \rightarrow B(r) \quad (16)$$

Therefore, we can generalize that for every red *nephew R(r)* that has a *B(g(r))*, the *R(r)* must become a black node, *B(r),* at the end of the 3R process, see Table 1.

Table 1. LR Case with an Inner Red Nephew to DB Node

| Steps | Structure & Case | Rotation | Rule Applied | | | Operated Nodes | Exempted Node | DB Removed | Tree Balanced |
|---|---|---|---|---|---|---|---|---|---|
| | | | $\Delta_{[DB,r,p]}$ | $\partial'_{[DB,p]}$ | $\partial''_{[r]}$ | | | | |
| 1. | LR case with an inner red nephew r of DB | Left-rotate s(DB) | - | - | - | s(DB) | - | No | No |
| 2. | LL case | - | Yes | - | - | DB, red inner nephew r, and p(DB) | - | Yes | No |
| 3. | LL case | Right-rotate new DB, remove DB | - | - | - | New DB | - | Yes | Yes |

Based on the foregoing analysis, we arrived at ***Corollary 3***.

▪ *Corollary 3:*

*In a RB tree, where there is **only** one red inner nephew to a DB with a black parent, then the case is a **LR** or **RL** case decided by the only red inner nephew; and as the **VIP** node, this nephew will have a color change.*

### 4.2.2 Right-Left Case of Double Black

***Example 2:*** *The case of a DB u with a black parent and one red inner nephew r.*

This Example 2 in Figure 7 is a **mirror** of the LR Case of Example 1 in Figure 6. As such, the same generalization that holds for the LR Case also hold for the RL case. Table 2 provides the highlight in this operations. In a list format, we state the different stages involved as

$$DB_{RL}^{B(p),InnerR(r)} = [DB, rightRotate(s(DB)), RR, \Delta_{[DB,r,p]}, leftRotate(newDB), remove\ DB] \quad (17)$$

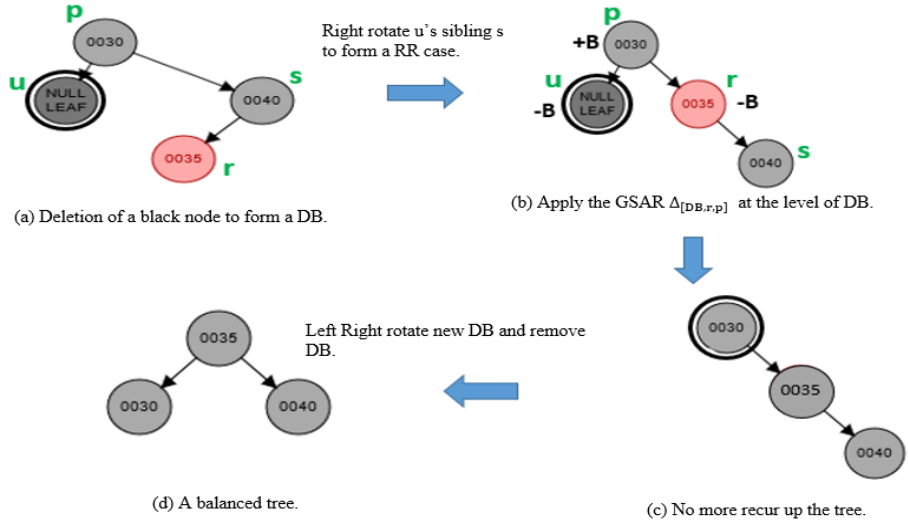

(a) Deletion of a black node to form a DB.

Right rotate u's sibling s to form a RR case.

(b) Apply the GSAR $\Delta_{[DB,r,p]}$ at the level of DB.

Left Right rotate new DB and remove DB.

(d) A balanced tree.

(c) No more recur up the tree.

Fig. 7. A RL case after the deletion of a black node.

Table 2. RL Case with an Inner Red Nephew to DB Node

| Steps | Structure & Case | Rotation | Rule Applied | | | Operated Nodes | Exempted Node | DB Removed | Tree Balanced |
|---|---|---|---|---|---|---|---|---|---|
| | | | $\Delta_{[DB,r,p]}$ | $\partial'_{[DB,p]}$ | $\partial''_{[r]}$ | | | | |
| 1. | RL case with an inner red nephew r of DB | Right-rotate s(DB) | - | - | - | s(DB) | - | No | No |
| 2. | RR case | - | Yes | - | - | DB, red inner nephew r, and p(DB) | - | Yes | No |
| 3. | RR case | Left-rotate new DB, remove DB | - | - | - | New DB | - | No | No |

Note that the given ***Corollary 3*** above also holds for the RL case of a **DB** with one red inner nephew ***r*** to a DB.

### 4.2.3 Left-Left Case with Two Red Nephews

**Example 3**: *The case DB u with a black parent and two red nephews r, and x.*

In Figure 8 we have a LL Case in which the DB has a black parent $B(p(40))$ and two red nephews which are the *red outer nephew **r*** given as ***R**(outerNephew(r(DB)))*, and the *red inner nephew **x*** as ***R**(innerNephew(x(DB)))* to ***DB***. Assigning values to parameters ***r*** and ***x***, we have, ***R**(outerNephew(r(20)))* and ***R**(innerNephew(x(35)))*. As mentioned earlier, since there are two red nephews ***r*** and ***x***, one of the two nephews must be the **VIP** *nephew*. Thus, the ***R**(outerNephew(r(20)))* becomes the **VIP** node. The node ***R**(outerNephew(r(20)))* is the *VIP node because i) it was used to determine the case as a LL case, and ii) changed its own color*. In Figure 8(a) we have a DB formed. In Figure 8(b) we right-rotate the black parent, ***p***, of **DB**, $B(p(40))$ to have BST (binary search tree). Then we apply the *partial rule1* $\partial'_{[DB,p]}$. *Partial rule1* operates on the **DB** and its *parent, **p***. Thus, we have ***DB – B = NULL_LEAF***, and $B(p(40))$ ***+ B = DB(40)***. This operations gives the tree a new ***DB(40)***. The ***new DB(40)*** is the original parent ***p*** of the original ***DB(u)***. The new ***DB(40)*** also has no place in a RB tree. So, to remove it, we recur up the RB tree and apply the *general rule* $\Delta_{[DB,r,p]}$ at the level of, *i)* the new ***DB(40)***, *ii)* ***R**(outerNephew(r(20)))* and their new black parent, $B(p(30))$ which was the original sibling *s(30)*, of the ***DB(u)***. From the *general rule* operation, we had, ***DB(40) - B = B(40)***, ***R**(outerNephew(r(20))) – B = B(20)* and ***B**(s(30)) + B = DB(30)*. Now, since the node ***DB(30)*** is a root node, every root is black in a RB tree. Table 3 presents these stages of operation in solving the 3R problem.

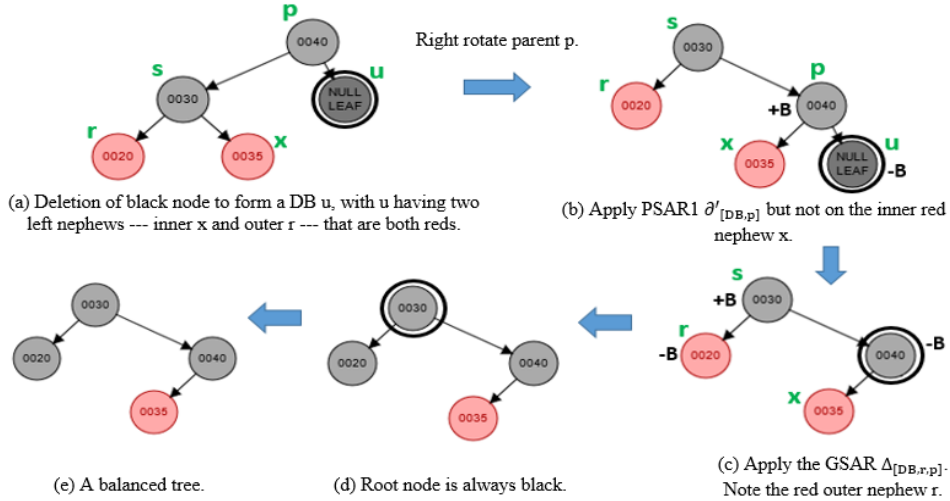

Fig. 8. LL case with two red nephews of the double black.

Table 3. LL Case with Two Red Nephews to DB Node

| Steps | Structure & Case | Rotation | Rule Applied | | | Operated Nodes | Exempted Node | DB Removed | Tree Balanced |
|---|---|---|---|---|---|---|---|---|---|
| | | | $\Delta_{[DB,r,p]}$ | $\partial'_{[DB,p]}$ | $\partial''_{[r]}$ | | | | |
| 1. | LL case with two red nephews of DB | Right-rotate p(DB) | - | - | - | p(DB) | - | No | No |
| 2. | LL case with two red nephews of DB | - | - | Yes | - | DB and p(DB) | Inner red nephew x | Yes | No |
| 3. | Binary search tree | - | Yes | - | - | New DB, outer nephew r, and new p(DB) | - | No | No |
| 4. | New DB is the root | - | - | - | - | Root is always black, then remove DB. | - | Yes | Yes |

In a list format, we show the series of operations and the different stages as

$$DB_{LL}^{B(p),TwoR(r,x)} = [DB, rightRotate(p), BST, \partial'_{[DB,p]}, newDB, \Delta_{[DB,r,p]}, newDB, newDB(root)] \qquad (18)$$

Using first-order logic formula *w.r.t.* the red **VIP** nephew **r** and the black grandparent of **r**, **B(g(r))** we state that

$$\forall g, r : B(g(r)) \land R(r) \rightarrow B(r) \qquad (19)$$

This LL Case example satisfies **Corollary 2** as earlier stated above.

#### 4.2.4 Right-Right Case with Two Red Nephews to DB

In this example, we provide two illustrative diagrams to demonstrate the SA algorithmic methods. While the first is a RR Case with two red nephews, the second is a RR case with one red nephew.

**Example 4**: *The case of DB u with a black parent and two red nephews x and r.*

Figure 9 presents a RR Case which is a **mirror** of the LL Case in Figure 8. As such, the same generalization holds for both structures. Table 4 depicts the series of operations involved, and the different stages of nodes' transformation before the tree was balanced as

$$DB_{RR}^{B(p),TwoR(r,x)} = [DB, leftRotate(p), BST, \partial'_{[DB,p]}, newDB, \Delta_{[DB,r,p]}, newDB(root), B(root)] \qquad (20)$$

Using first-order logic formula *w.r.t.* the red **VIP** nephew **r** and black grandparent of **r**, **B(g(r))**, we state that

$$\forall g, r : B(g(r)) \land R(r) \rightarrow B(r) \qquad (21)$$

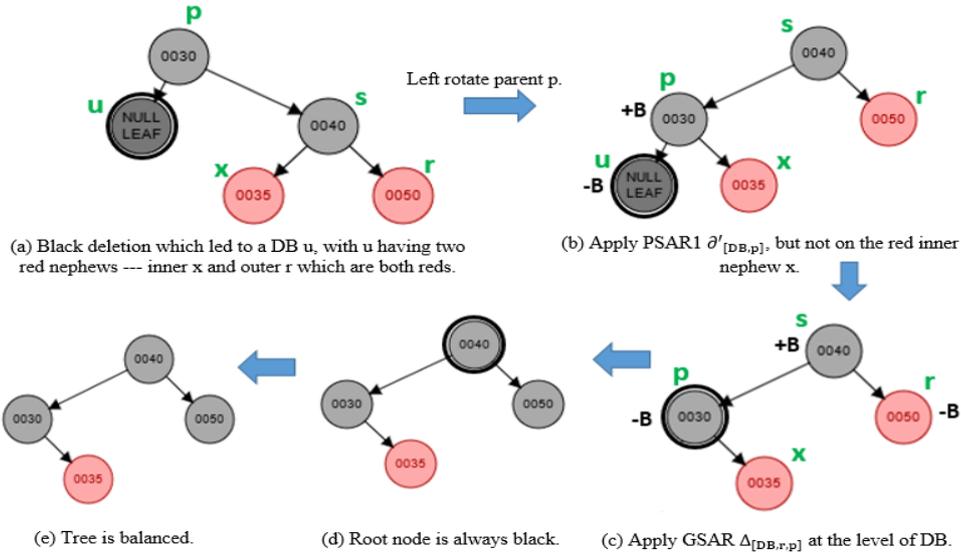

Fig. 9. RR case with two red nephews of the double black node.

Table 4. RR Case with Two Red Nephews to DB Node

| Steps | Structure & Case | Rotation | Rule Applied | | | Operated Nodes | Exempted Node | DB Removed | Tree Balanced |
|---|---|---|---|---|---|---|---|---|---|
| | | | $\Delta_{[DB,r,p]}$ | $\partial'_{[DB,p]}$ | $\partial''_{[r]}$ | | | | |
| 1. | RR case with two red nephews of DB | Left-rotate p(DB) | - | - | - | p(DB) | - | No | No |
| 2. | Binary search tree | - | - | Yes | - | DB and p(DB) | Inner red nephew x | No | No |
| 3. | Binary search tree | - | Yes | - | - | New DB, outer nephew r, and new p(DB) | - | No | No |
| 4. | New DB is the root | - | - | - | - | Root is always black. Remove DB. | - | Yes | Yes |

This RR Case example satisfies **Corollary 2** as earlier stated above.

**Example 5:** *The RR Case of DB u with a black parent and one red outer nephew r.*

When compared to the RR Case in the preceding Example 4 (Figure 9), Example 5 (Figure 10) has only one *red outer nephew r*, given as $R(outerNephew(r(DB)))$, to **DB**. This means that Example 5 is a similar case to the foregoing case in Example 4. Though in Example 5, the tree has no inner nephew node but their similarity lies in the **VIP** *nephew r* that decided the case as well as taking part in the color change operation of the SA algebraic rules. *Suppose* there was an *red inner nephew* in Example 5, it will not play any part in the color change operation. After balancing the tree, Example 4 and Example 5 have equal number of *blackHeight = 2* and thus the same generalization. Table 5 summarizes the steps invloved.

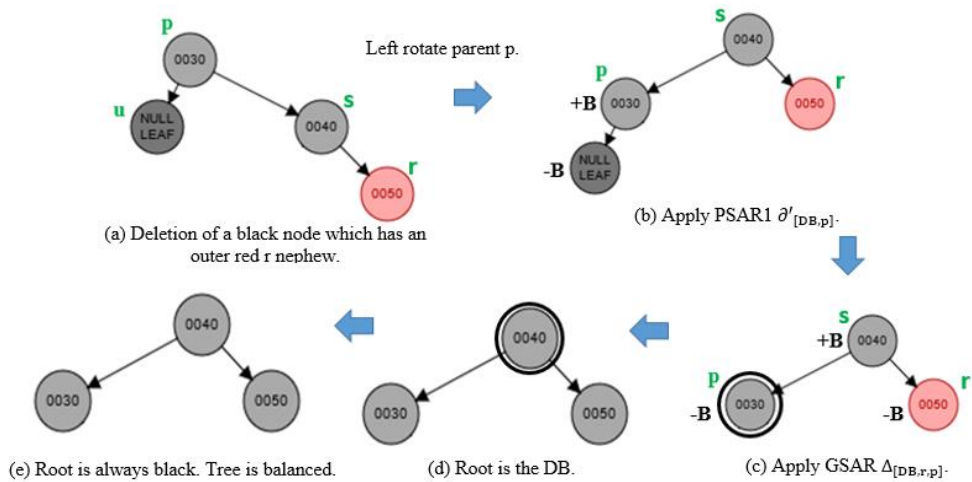

Fig. 10. RR case with one red outer nephew to the double-black node.

Table 5. RR Case with One Red Outer Nephews to DB Node

| Steps | Structure & Case | Rotation | Rule Applied | | | Operated Nodes | Exempted Node | DB Removed | Tree Balanced |
|---|---|---|---|---|---|---|---|---|---|
| | | | $\Delta_{[DB,r,p]}$ | $\partial'_{[DB,p]}$ | $\partial''_{[r]}$ | | | | |
| 1. | RR case with one outer red nephews of DB | Left-rotate p(DB) | - | - | - | p(DB) | - | No | No |
| 2. | Binary search tree | - | - | Yes | - | DB and p(DB) | - | No | No |
| 3. | Binary search tree | - | Yes | - | - | New DB, outer nephew r, and new p(DB) | - | No | No |
| 4. | New DB is the root | - | - | - | - | Root is always black. Remove DB. | - | Yes | Yes |

Based on the foregoing analysis, we therefore arrived at ***Corollary 4***.

- ***Corollary 4:***

In a RB tree, if there is **only** one red outer nephew to a DB with a black parent, then the case is a ***LL*** or ***RR*** case decided by the only red outer nephew; and as the ***VIP*** node inclusively, this nephew will have a color change.

### 4.2.5 Left-Left Case with Two Black Nephews

***Example 6***: The case of DB *u* with a black parent and two black nephews *r*, and *x*.

In this demo, we present a LL Case, with a black parent to DB, ***B(p(DB))***, and two black nephews ***r*** and ***x***, namely, the *black outer nephew **x***, ***B(outerNephew(x(DB)))***, and *black inner nephew **r***, ***B(innerNephew(r(DB)))*** in Figure 11. Assigning values to ***r*** and ***x***, we have ***B(outerNephew(x(20)))*** and ***B(innerNephew(r(35)))***. With two black nephews to the DB, the **VIP** *nephew* becomes the ***B(innerNephew(r(35)))***. *It is the **VIP** node **because** it is the nephew **r** that had a color change but **did not determine** the case as an LL to have the tree balanced*. In Figure 11(a) a **DB** was formed. In Figure 11(b) we right-rotate the ***B(p(DB))***. Then in Figure 11(c) we applied the *general rule* $\Delta_{[DB,r,p]}$ to i) ***DB*** node, ii) the ***B(innerNephew(r(DB)))***; and iii) the ***B(p(DB))***. by evaluation, ***DB − B = NULL_LEAF***, ***B(innerNephew(r(35))) − B = R(r(35))***, and ***B(p(40)) + B = DB(p(40))***. This operation led to a ***new DB*** formation i.e. ***DB(40)***. Note that the **VIP** nephew has changed color from black ***B(r(35))*** to red ***R(r(35))***. With a ***new DB(40)***, we recur up the tree to remove the ***new DB(40)***. At this point we invoked the *partial rule1* $\partial'_{[DB,p]}$ on the ***DB(40)*** and its black ***new*** parent, ***B(p(30))*** (which was the *original black sibling **s**, *$_{original}$***B(s(DB))*** of DB). By another evaluation, we had ***newDB(40) − B = B(40)*** and $_{original}$***B(s(30)) + B = newDB(30)*** that took us to another ***new DB(30)***. However, since the ***newDB(30)*** is a root node, according to the properties of the RB tree, every root node is black. So, in the end we had a black root node, ***B(root(30))*** in Figure 11(e) to have a balanced tree. Table 6 summarizes the operations folowed in resolving the 3R problem. In a list, we also represent these operations and the different stages as

$$DB_{LL}^{B(p),TwoB(r,x)} = [DB, rightRotate(p), BST, \Delta_{[DB,r,p]}, newDB, \partial'_{[DB,p]}, newDB, newDB(root), B(root)] \quad (22)$$

Using first order logic formula *w.r.t.* to the **VIP** nephew *r* and *g(r)*, we state that

$$\forall g, r : B(g(r)) \wedge B(r) \rightarrow {\color{red}R}(r) \quad (23)$$

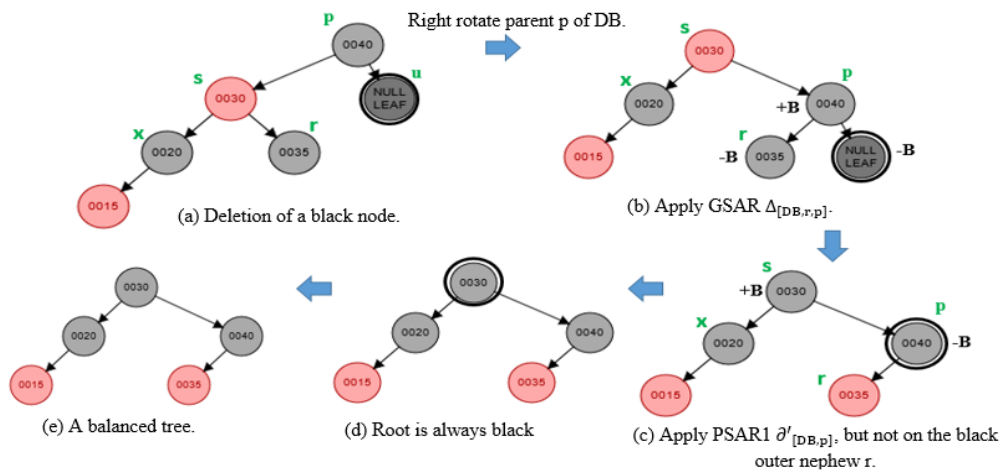

Fig. 11. LL case with two black nephews of double-black.

Table 6. LL Case with Two Black Nephews to DB Node

| Steps | Structure & Case | Rotation | Rule Applied | | | Operated Nodes | Exempted Node | DB Removed | Tree Balanced |
|---|---|---|---|---|---|---|---|---|---|
| | | | $\Delta_{[DB,r,p]}$ | $\partial'_{[DB,p]}$ | $\partial''_{[r]}$ | | | | |
| 1. | LL case with two black nephews of DB | Right-rotate p(DB) | - | - | - | p(DB) | - | No | No |
| 2. | Binary search tree | - | Yes | - | - | DB(u), black inner nephew r, and p(DB) | - | No | No |
| 3. | Binary search tree | - | - | Yes | - | New DB, and new p(DB) i.e. old s(DB) | Black outer nephew x | No | No |
| 4. | New DB is the root | - | - | - | - | Root is always black, then remove DB. | - | Yes | Yes |

Based on the foregoing analysis, we arrived at ***Corollary 5***.

▪ ***Corollary 5:***

*In a RB tree, if there are two black nephews to a DB with a black parent, the case is a **LL** or **RR** case decided by the **black outer nephew** but the **VIP** node would be the **black inner nephew** that will have a color change.*

Note that from **Examples 1 - 5 (Figures 6 – 10)**, the **VIP** nephew node *r* has always performed two functions, which are, i) deciding a rotation case and changing its colors. However, that changed in Example 6 (Figure 11), where we found an **exemption**. Instead, while the *black outer nephew **x*** was the node that **determined** the case, the *black inner nephew **r*** became the node that had a change of **color** to balance *black heights* in the tree.

### 4.2.6 Right-Right Case with Two Black Nephews

***Example 7:*** *The case of DB u with a black parent and two black nephews r, and x.*

Figure 12 is a RR Case which is a ***mirror*** of the LL Case in Example 6 (Figure 11). As such, the same generalization holds for both examples. Table 7 depicts the series of operations involved. In a list format, we represent these operations and the different stages as

$$DB_{RR}^{B(p),TwoB(r,x)} = [DB, leftRotate(p), BST, \Delta_{[DB,r,p]}, \partial'_{[DB,p]}, B(root)] \tag{24}$$

The first-order logic formula w.r.t. the black **VIP** nephew *r* and the black grandparent of *r*, ***B(g(r))*** states that

$$\forall g, r : B(g(r)) \wedge B(r) \rightarrow R(r) \tag{25}$$

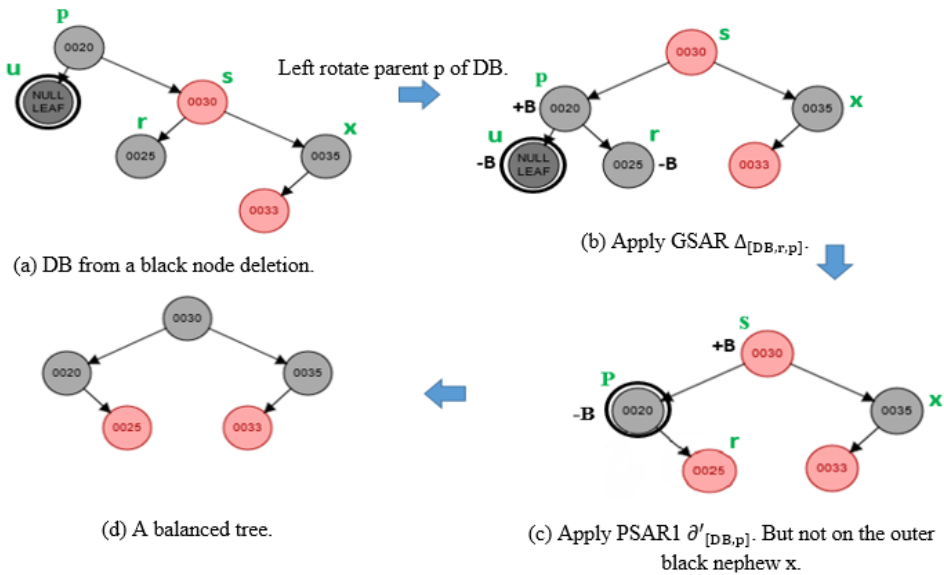

(a) DB from a black node deletion.

Left rotate parent p of DB.

(b) Apply GSAR $\Delta_{[DB,r,p]}$.

(c) Apply PSAR1 $\partial'_{[DB,p]}$. But not on the outer black nephew x.

(d) A balanced tree.

Fig. 12. RR case with two black nephews of the double black node.

Table 7. RR Case with Two Black Nephews to DB Node

| Steps | Structure & Case | Rotation | Rule Applied | | | Operated Nodes | Exempted Node | DB Removed | Tree Balanced |
|---|---|---|---|---|---|---|---|---|---|
| | | | $\Delta_{[DB,r,p]}$ | $\partial'_{[DB,p]}$ | $\partial''_{[r]}$ | | | | |
| 1. | RR case with two black nephews of DB | Left-rotate p(DB) | - | - | - | p(DB) | - | No | No |
| 2. | Binary search tree | - | Yes | - | - | DB(u), inner black nephew r, and p(DB) | - | No | No |
| 3. | Binary search tree | - | - | Yes | - | New DB, and new p(DB) i.e. old s(DB) | Outer black nephew x | No | No |
| 4. | New DB is the root | - | - | - | - | Root is always black. Remove DB. | - | Yes | Yes |

This RR Case example satisfies **Corollary 5** as stated above.

### 4.3 PART II: Examples of Removing Double-Black with a Red Parent

Having taken several Case examples with black parents of DB, **B(p(DB))**; now we consider the Cases of red parent, **R(p(DB))** to DB. In the following examples, we shall consider the cases in Examples 8 - 10 (Figures 13 -15).

#### 4.3.1 Left-Right Case with one Red Nephew

**Example 8**: *The case of DB u with a red parent and one red inner nephew r.*

Figure 13 presents our first case of a red parent **p** to DB, **R(p(DB))**, in a LR Case. In this case, our **VIP** node becomes the *red inner nephew **r*** to the DB node, which is symbolized as **R(innerNephew(r(DB)))**. As the only nephew node, **r**, to the DB, the **R(**innerNephew**(r(DB)))** thus determines the case to be a LR case. To resolve the 3R problem of DB removal, we left-rotate the black sibling **s** of DB, **B(s(DB))**, to take us to a LL case, *Figure 13(b)*. Whilst the nodes still keep their color, we apply the *general rule* $\Delta_{[DB,r,p]}$ at the level of DB such that **DB - B = NULL_LEAF, R(**innerNephew**(r(19))) - B = B(19)**, and **R(p(20)) + B = B(p(20))**, *Figure 13(c)*. Still, as a LL case, we right-rotate the **B(p(20))**, *Figure 13(d)*. Then apply the *partial rule2,* $\partial''_{[r]}$, to the black **VIP** *nephew **r***, and the tree is balanced, *Figure 13(e).* The highlight of this procedure is given in Table 8. In a list format, we represent the different operations and stages as

$$DB_{LR}^{R(p),InnerR(r)} = [DB, leftRotate(s), LL, \Delta_{[DB,r,p]}, rightRotate(p), BST, \partial''_{[r]}] \qquad (26)$$

In first-order logic formula *w.r.t.* the red **VIP** *nephew **r*** at the point of formation of the **DB**, and the ***red grandparent of r, R(g(r))***, we conclude that

$$\forall g, r : R(g(r)) \wedge (R(r) \rightarrow B(r) \wedge B(r) \rightarrow R(r)) \qquad (27)$$

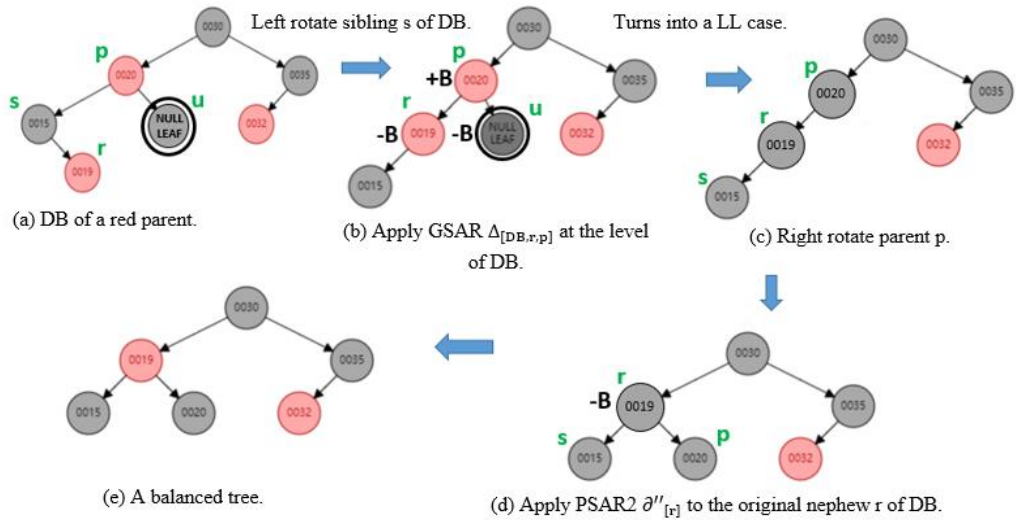

(a) DB of a red parent.

(b) Apply GSAR $\Delta_{[DB,r,p]}$ at the level of DB.

(c) Right rotate parent p.

(e) A balanced tree.

(d) Apply PSAR2 $\partial''_{[r]}$ to the original nephew r of DB.

Fig. 13. LR case with one red inner nephew to double-black.

Table 8. LR Case with An Inner Red Nephews to DB Node

| Steps | Structure & Case | Rotation | Rule Applied | | | Operated Nodes | Exempted Node | DB Removed | Tree Balanced |
|---|---|---|---|---|---|---|---|---|---|
| | | | $\Delta_{[DB,r,p]}$ | $\partial'_{[DB,p]}$ | $\partial''_{[r]}$ | | | | |
| 1. | LR case with one inner red nephew of DB | Left-rotate s(DB) | - | - | - | s(DB) | - | No | No |
| 2. | LL case | - | Yes | - | - | DB(u), inner red nephew r, and p(DB) | - | Yes | No |
| 3. | LL case | Right-rotate original p(DB) | - | - | - | p(DB) | - | Yes | No |
| 4. | Binary search tree | - | - | - | Yes | Original nephew r(DB) | r's two new children | Yes | Yes |

Note how the **VIP** *nephew r* changed its color as shown in the logic formula, firstly, from red $R(r)$ to black $B(r)$ and then from black $B(r)$ to red $R(r)$. Based on the foregoing analysis, we arrived at **Corollary 6**.

▪ *Corollary 6:*

*In a RB tree, if there is **only** one red inner nephew to a DB with a red parent, then the case is a **LR** or **RL** case decided by the only red inner nephew; and as the **VIP** node, this nephew will have a color change.*

### 4.3.2 Right-Left Case with one Red Nephew

***Example 9:*** *The case of DB u with a red parent and one red inner nephew r.*

Figure 14 of ***Example 9*** is a RL Case which *mirrors* the LR case of ***Example 8*** (Figure 13). Therefore, the same generalization holds for both Examples 8 and 9. The highlight of the procedure is also shown in Table 9. In a list format, we state the different stages involved as

$$DB_{RL}^{R(p),InnerR(r)} = [DB, rightRotate(s), RR, \Delta_{[DB,r,p]}, leftRotate(p), BST, \partial''_{[r]}] \qquad (28)$$

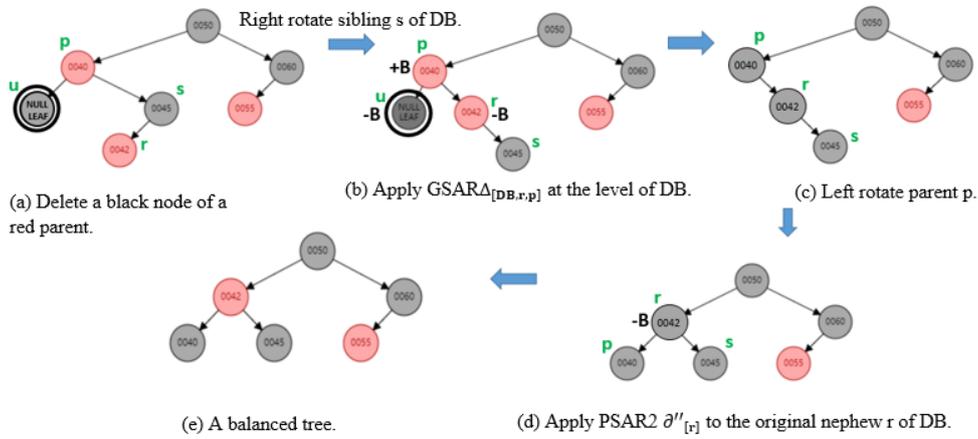

(a) Delete a black node of a red parent.

(b) Apply GSAR $\Delta_{[DB,r,p]}$ at the level of DB.

(c) Left rotate parent p.

(d) Apply PSAR2 $\partial''_{[r]}$ to the original nephew r of DB.

(e) A balanced tree.

Fig. 14. RL case with one red inner nephew to double-black.

Table 9. RL Case with An Inner Red Nephews to DB Node

| Steps | Structure & Case | Rotation | Rule Applied | | | Operated Nodes | Exempted Node | DB Removed | Tree Balanced |
|---|---|---|---|---|---|---|---|---|---|
| | | | $\Delta_{[DB,r,p]}$ | $\partial'_{[DB,p]}$ | $\partial''_{[r]}$ | | | | |
| 1. | RL case with one inner red nephew of DB | Right-rotate s(DB) | - | - | - | s(DB) | - | No | No |
| 2. | RR case | - | Yes | - | - | DB, inner red nephew r, and p(DB) | - | Yes | No |
| 3. | RR case | Left-rotate original p(DB) | - | - | - | p(DB) | - | Yes | No |
| 4. | Binary search tree | - | - | - | Yes | Original nephew r(DB) | r's two new children | Yes | Yes |

This RL Case example satisfies ***Corollary 6*** as stated above.

### 4.3.3  Left-Left Case with one Red Nephew

**Example 10**: *The case of a DB u with a red parent and one red outer nephew r.*

In Figure 15 we have a LL Case. The LL case has a red parent, *p*, **R(p(DB))**, a black sibling, *s*, **B(s(DB))**, and one *red outer nephew **r**,* **R(outerNephew(r(DB)))**, to DB, respectively. The **R(out**erNephew(r(DB))) is the only nephew node and it doubles as the *VIP node,* and as the node that must have a *color change* at the end of the 3R process. To resolve the 3R problem, we right-rotate **R(p(30))** to have a RR Case, *Figure 15(b)*, and then applied the *partial rule1* $\partial'_{[DB,p]}$ at the level of the DB such that **DB - B = NULL_LEAF**, and **R(p(30)) + B = B(30)**, *Figure 15(c)*. Then, called the *partial rule2* $\partial''_{[r]}$ on the **R(outerNephew(r(15)))** where **R(outerNephew(r(15))) - B = B(15)** to have a balanced RB tree, *Figure 15(d)*. The highlight of this procedure is given in Table 10. In a list format, the different stage of this SA operation is stated as

$$DB_{LL}^{R(p),OuterR(r)} = [DB, rightRotate(p), RR, \partial'_{[DB,p]}, BST, \partial''_{[r]}] \tag{29}$$

The first-order logic representation *w.r.t.* the red VIP *nephew r* and the red *grandparent* of *r*, **R(g(r))** is given as

$$\forall g, r : R(g(r)) \wedge R(r) \rightarrow B(r) \tag{30}$$

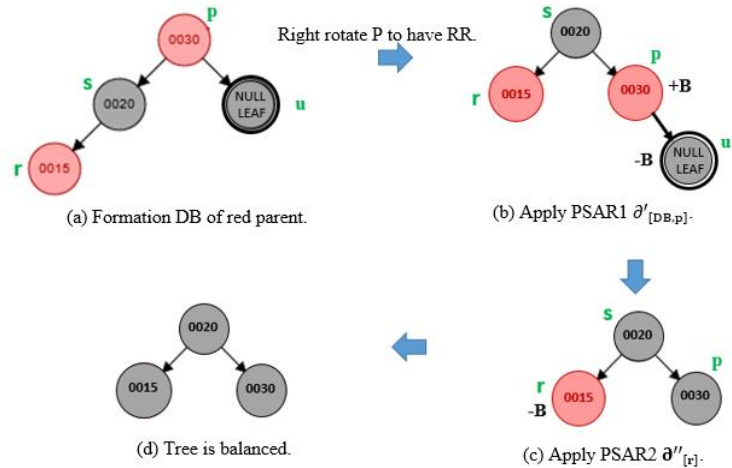

(a) Formation DB of red parent.

Right rotate P to have RR.

(b) Apply PSAR1 $\partial'_{[DB,p]}$.

(c) Apply PSAR2 $\partial''_{[r]}$.

(d) Tree is balanced.

Fig. 15. LL case with one red outer nephew to double-black.

Table 10. LL Case with an Outer Red Nephew to DB Node

| Steps | Structure & Case | Rotation | Rule Applied | | | Operated Nodes | Exempted Node | DB Removed | Tree Balanced |
|---|---|---|---|---|---|---|---|---|---|
| | | | $\Delta_{[DB,r,p]}$ | $\partial'_{[DB,p]}$ | $\partial''_{[r]}$ | | | | |
| 1. | LL case with one outer red nephew of DB | Right-rotate p(DB) | - | - | - | p(DB) | - | No | No |
| 2. | RR case | - | - | Yes | - | DB, outer red nephew r, R(r(DB)) | - | Yes | No |
| 3. | BST | - | - | - | Yes | Outer red nephew R(r(DB)) | Original p(DB) & s(DB) | Yes | Yes |

Based on the foregoing analysis, we arrived **Collary 7**:

▪ *Collary 7:*

*In a RB tree, if there is **only** one red outer nephew to a DB with a red parent, then the case is a **LL or RR** case decided by the only red outer nephew; and as the **VIP** node, this nephew will have a color change.*

The **mirror** of this LL case example is the RR Case. Thus, the same generalization for the LL Case also holds for its counterpart RR Case. In a list format, the different stages of the RR case operation is given as

$$DB_{RR}^{R(p),OuterR(r)} = [DB, leftRotate(p), LL, \partial'_{[DB,p]}, BST, \partial''_{[r]}] \tag{31}$$

This LL Case example satisfies **Corollary 7** as stated above.

*4.4 Algorithm*

This is the algorithm of *symbolic-algebraic arithmetic* (SA) operation for DB *removal* with *rotation*, and *recoloring* (*3Rs*) of nodes in RB trees.

---

**Algorithm of Double_Black with Rotation Removal by Algebraic Operation**

DoubleBlackWithRotation_Removal(T, p, v, u, r, s, x)
1. **IF** root = DB **THEN**
2.     DB(root) - B => B(root)
3. **ENDIF**
4. **IF** root = x && deleted(node) == x **THEN**
5.     color(x) = B
6. **ENDIF**
7. **IF** color(x) = R
8.     **IF** deleted(node) == x **THEN**
9.         R(x) + B => B(x)
10.     **ENDIF**
11. **ENDIF**
12. **IF** color(x) = B
13.     **IF** deleted(node) == x **THEN**
14.         B(x) + B => DB(u)
15.         **WHILE** DB(u):
16.             /** LR case */
17.             **IF** p(DB(u)) == R && DB(u(hasInnerRedNephew(r))) && LR_case **THEN**
18.                 method_LR_Rotation(p,u,s,r)              //LR Call to line 88
19.                 apply_GSAR()                             //GSAR rule call to line 94
20.                 method_LL_Rotation(p,u,s,r)              //LL Call to line 103
21.                 apply_PSAR2()                            //PSAR2 rule call to line 109
22.             **ENDIF**
23.             /** RL case */
24.             **IF** p(DB(u)) == R && DB(u(hasInnerRedNephew(r))) && RL_case **THEN**
25.                 method_RL_Rotation(p, u, s, r)   //RL call to line 91
26.                 apply_GSAR()                             //GSAR rule call to line 94
27.                 method_RR_Rotation(p,u,s,r)       //RR call to line 106
28.                 apply_PSAR2()                            //PSAR2 rule call to line 109
29.             **ENDIF**
30.             /* LR to LL case */
31.             **IF** p(DB(u)) == B && DB(u(hasInnerRedNephew(r))) && LR_case **THEN**
32.                 method_LR_Rotation(p,u,s,r)              //LR Call to line 88
33.                 apply_GSAR()                             //LL Case: GSAR rule call to line 94
34.                 method_LL_Rotation(p,u,s,r)              //LL Call to line 103
35.                 newDB(P) - B => B(p)                     //Tree balanced
36.             **ELSE** // RL to RR case
37.                 method_RL_Rotation(p,u,s,r)              //RL call to line 91
38.                 apply_GSAR();                            //RR Case: GSAR rule call to line 94
39.                 method_RR_Rotation(p,u,s,r)              //RR Case
40.                 newDB(P) - B => B(p)                     //Tree balanced
41.             **ENDIF**
42.             /* LL or RR case*/
43.             **IF** p(DB(u)) == B && DB(u(hasInnerRedNephew(x)) && (hasOuterRedNephew(r))) **THEN**
44.                 //LL
45.                 method_LL_Rotation(p,u,s,r)      //LL Call to line 103
46.                 apply_PSAR1()                            // PSAR1 rule call to line 112
47.                 apply_GSAR()                             //LL Case: GSAR rule call to line 94
48.             **ELSE** //RR
49.                 method_RR_Rotation(p,u,s,r)  //RR Call to line 106
50.                 apply_PSAR1()                            // PSAR1 rule call to line 112
51.                 apply_GSAR()                             //RR Case: GSAR rule call to line 94
52.             **ENDIF**
53.             /* LL or RR case*/
54.             **IF** p(DB(u)) == B && DB(u(hasOutRedNephew(r))) **THEN**
55.                 method_LL_Rotation(p,u,s,r)   //LL Call to line 103
56.                 apply_PSAR1()                            // PSAR1 rule call to line 112
57.                 apply_GSAR()                             //LL Case: GSAR rule call to line 94
58.             **ELSE**
59.                 method_RR_Rotation(p,u,s,r)   //RR Call to line 106
60.                 apply_PSAR1()                            // PSAR1 rule call to line 109
61.                 apply_GSAR()                             //RR Case: GSAR rule call to line 94
62.             **ENDIF**
63.             /** LL or RR case */
64.             **IF** p(DB(u)) == B && DB(u(hasInnerBlackNephew(r)) && (hasOuterBlackNephew(x))) && LL_Case **THEN**
65.                 method_LL_Rotation(p,u,s,r)  //LL Call to line 103
66.                 apply_GSAR()                             //LL Case: GSAR rule call to line 94
67.                 apply_PSAR1()                            //PSAR1 rule call to line 112
68.             **ELSE**

---

```
69.                 method_RR_Rotation(p,u,s,r)//RR call to line 106
70.                 apply_GSAR()                        //RR Case: GSAR rule call to line 94
71.                 apply_PSAR1()                       //PSAR1 rule call to line 112
72.             ENDIF
73.         IF p(DB(u)) == R && DB(u(hasOuterRedNephew(r))) && LL_case THEN
74.             method_LL_Rotation(p,u,s,r)
75.             apply_PSAR1()
76.             apply_PSAR2()
77.         ELSE
78.             method_RR_Rotation(p,u,s,r)
79.             apply_PSAR1()
80.             apply_PSAR2()
81.         ENDIF
82.       ENDWHILE
83.     ENDIF
84. ENDIF
85. Output Tree
86.
87.     /** Methods Definition */
88.     method_LR_Rotation(p,u,s,r){
89.             leftRotate(DB(u));                      //LL case 3
90.     }
91.     method_RL_Rotation(p,u,s,r){
92.             rightRotate(s(DB(u)));                  //RR case 3
93.     }
94.     apply_GSAR(p,u,s,r){
95.             DB(u) - B => B(x);
96.             R(r) - B => B(r);
97.             IF p(DB(x)) = R
98.                     R(p(DB(x))) + B => B(p);
99.             ELSE
100.                    B(P(DB(x))) + B => newDB(u)
101.            // recur up the tree
102.     }
103.     method_LL_Rotation(p,u,s,r){
104.            rightRotate(B(p));                       //original parent of DB
105.     }
106.     method_RR_Rotation(p,u,s,r){
107.            leftRotate(B(p));                        //original parent of DB
108.     }
109.     apply_PSAR2(){
110.            B(r) - B => R(r)                         //Tree is balanced
111.     }
112.     apply_PSAR1(){
113.            DB(u) - B => B(NULL_LEAF)    //B(NULL_LEAF) = DB(u) - B
114.            B(p(DB(u))) + B => DB(p)        // DB(P) = B(p(DB(u))) + B
115.     }
116. END PROCEDURE
```

### 4.5 Findings

Rotations in binary search trees such as AVL and RB trees is part of a tree balancing operation that leads to a balanced state of the trees. In this study, all the structural cases of **DB** removal that involved rotation; we found that *rotation* is the first operation that occurs before the application of node color-change by the SA algorithms. Having looked at the evidence of the SA algorithmic operations on subparts of the RB tree; in Examples 11 − 13, we look at the behavior of SA algorithms on RB trees with bigger depths and nodes using tree diagrams.

***Example 11:*** *Pictorial Analysis of a RL Case.*

This example in Figure 16 is a RL Case denoted as $DB_{RL}^{B(p),InnerR(r)}$ given in eq. (28). This Case is similar to the example in Figure 14 above.

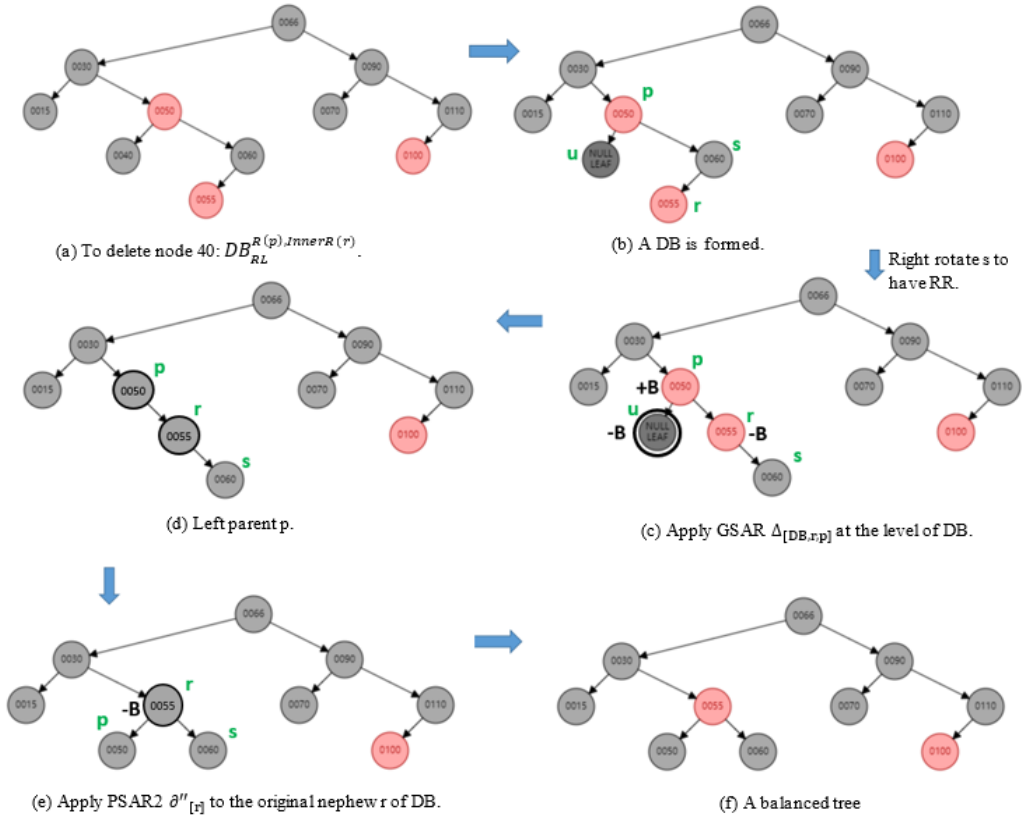

(a) To delete node 40: $DB_{RL}^{R(p),InnerR(r)}$.

(b) A DB is formed.

(c) Apply GSAR $\Delta_{[DB,r,p]}$ at the level of DB.

Right rotates s to have RR.

(d) Left parent p.

(e) Apply PSAR2 $\partial''_{[r]}$ to the original nephew r of DB.

(f) A balanced tree

Fig. 16. RL case with one red inner nephew to double-black.

***Example 12:*** *Pictorial Analysis of a LL Case.*

This pictorial analysis in Figure 17 is a complex LL Case as denoted by $DB_{LL}^{B(p),TwoB(r,x)}$ in eq. (18). We deleted node 70 in Figure 17(a) to form a DB. Then right-rotate the ***B(p(DB))*** i.e. ***B(p(66))*** to have a RL Case, Figure 17(b). Apply the *general rule* $\Delta_{[DB,r,p]}$ to remove the DB. At this point, a ***new DB(66)*** is formed in Figure 17(c) to have a RL Case. As we recur up the tree to balance the tree; we right-rotate the new ***DB(66)*** to have a BST; and finally applied the *partial rule2,* $\partial''_{[r]}$. Enumerating the different stages and operations, we have

$$DB_{LL}^{B(p),TwoB(r,x)} = [DB, rightRotate(p), \Delta_{[DB,r,p]}, RL, \Delta_{[DB,r,p]}, BST, \partial''_{[r]}] \tag{32}$$

This case example is also a **special case** like that of Example 7 Figure 12 $DB_{RR}^{B(p),TwoB(r,x)}$, see eq. (24), where the Case was not determined by the **VIP** *nephew r* but by the *outer black nephew x* whose node color does not change. Removing the DB in this case and balancing the tree, involved applying the *general rule* $\Delta_{[DB,r,p]}$ twice and then the *partial rule2* $\partial''_{[r]}$.

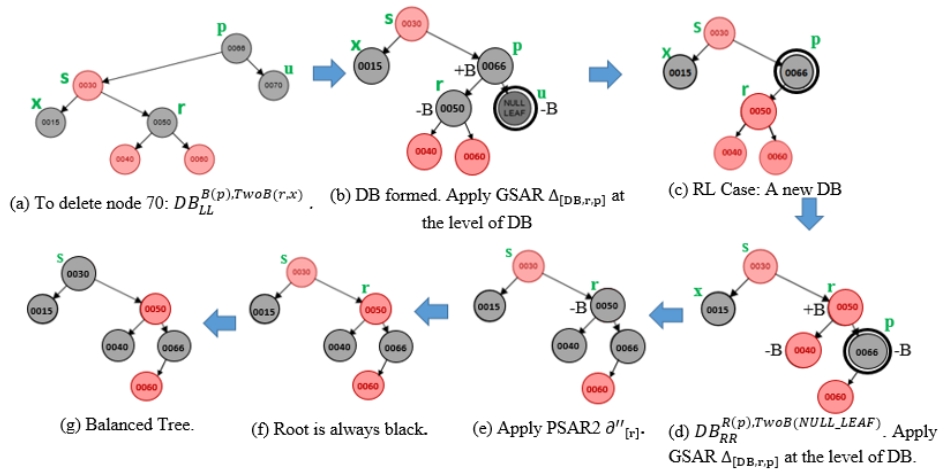

(a) To delete node 70: $DB_{LL}^{B(p),TwoB(r,x)}$.

(b) DB formed. Apply GSAR $\Delta_{[DB,r,p]}$ at the level of DB

(c) RL Case: A new DB

(d) $DB_{RR}^{R(p),TwoB(NULL\_LEAF)}$. Apply GSAR $\Delta_{[DB,r,p]}$ at the level of DB.

(e) Apply PSAR2 $\partial''_{[r]}$

(f) Root is always black.

(g) Balanced Tree.

Fig. 17. Complex LL case with two black nephews to double-black.

***Example 13:*** *Pictorial Analysis of a RL Case.*

In Figure 18 is a RL case that fits into the expression $DB_{RL}^{B(p),InnerR(r)}$ in eq. (28). As demonstrated in Figure 7 and its mirror case in Figure 6; the *inner red nephew **r*** is the **VIP** node that determined the case as a RL. To balance this tree, only the *general rule* $\Delta_{[DB,r,p]}$ is applicable.

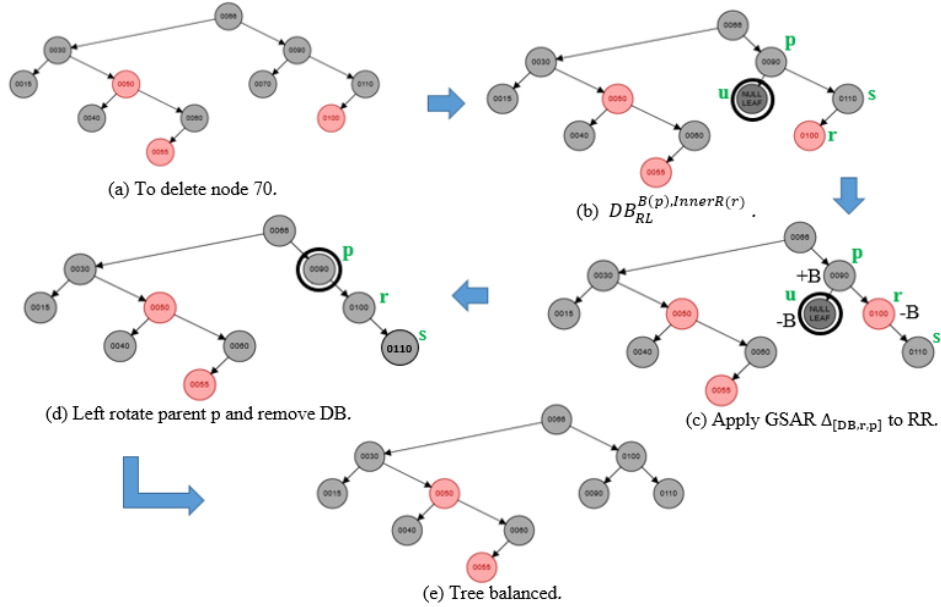

(a) To delete node 70.

(b) $DB_{RL}^{B(p),InnerR(r)}$.

(c) Apply GSAR $\Delta_{[DB,r,p]}$ to RR.

(d) Left rotate parent p and remove DB.

(e) Tree balanced.

Fig. 18. RL case with one red inner nephew to double-black.

### 4.5.1 Sibling Node s versus Nephew node r

On a final note, recall that our initial study in [4] worked on DB removal without rotation of nodes. In the aforementioned work, where the *general SA rule* was first projected, we applied the rule to nodes ***[DB, s, p]***. Because there was no rotation, the SA operated on the ***DB*** node, its sibling *s*, and its parent ***p*** only. In this paper which deals with node rotation, we realized that **"no one rule fits all"**. This is because of the complexity of handling DB removal with rotation. To handle this complexity, we needed to refine the *general SA rule* to derive two additional rule, namely, the *partial rule1* and *partial rule2*.

Rotation results in the change of position in a tree. Now, let's denote a ***new sibling*** of **DB** as *s\**. Depending on the case, for instance, LR or RL Case; when a DB is rotated, a DB gets a *new sibling **s\***, which is the original nephew *r* to DB. See the examples in Figure 6(b), 7(b), 8(c), 9(c), 10(c), 11(b), 12(b), 13(b), 14(b), 16(c), 17(b), and 18(c). The *general SA rule* which is always operated on three connected nodes is operated on the ***[DB, r, p]*** after rotation, where ***r*** becomes the newly adopted sibling *s\** to **DB**. In a nutshell, since an original nephew such as *r* has transformed from a *nephew* to the *sibling **s\***; then we state that the $\Delta_{[DB,r,p]} = \Delta_{[DB,s^*,p]}$ as a result of node rotation. Except for Example 10 Figure 15 expression $DB_{LL}^{R(p),OuterR(r)}$, eq. 31; all the other examples have in one way or the other used the $\Delta_{[DB,r,p]}$ rule to balance a RB tree. In this LL Case of Example 10, the DB removal problem combined two partial rules: *partial 1,* $\partial'_{[DB,p]}$ and *partial 2,* $\partial''_{[r]}$. Therefore, we conclude that

$$\Delta_{[DB,r,p]} = \partial'_{[DB,p]} + \partial''_{[r]} \tag{33}$$

$$\partial'_{[DB,p]} = \Delta_{[DB,r,p]} - \partial''_{[r]} \tag{34}$$

$$\partial''_{[r]} = \Delta_{[DB,r,p]} - \partial'_{[DB,p]} \tag{35}$$

### 4.6.1 First Comparative Example of the Traditional RB Algorithm vs. SA Algorithm:

Table 10 compares the performance of the (Conventional) Traditional RB algorithm (TA) against our SA algorithm. As presented by the number of steps in Table 10, both TA and SA methods have shown equal performance based on the same number of steps or runtime of 5 each to balance the given RB tree.

Table 10. Side-by-Side Performance Comparison of the TA vs. SA with Equal Number of Steps.

| Steps | Traditional Algorithm | Symbolic-Arithmetic Algorithm |
|---|---|---|
| 1. | Delete 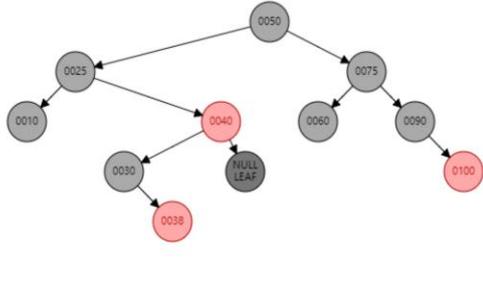 | Delete 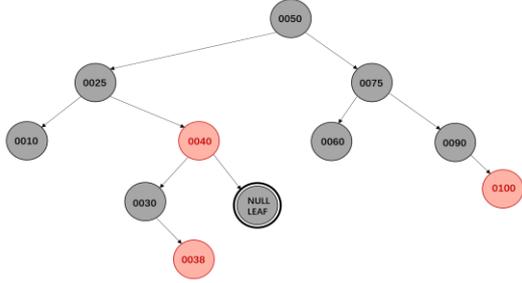 |
| 2. | Rotate 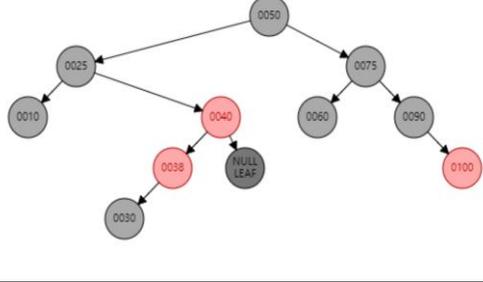 | Rotate 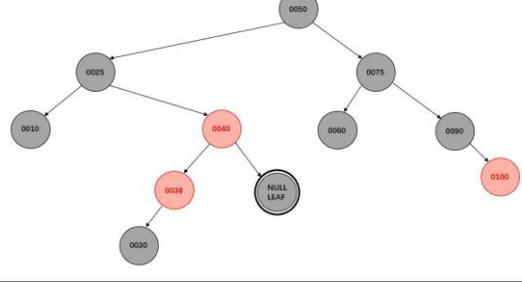 |
| 3. | Re-color 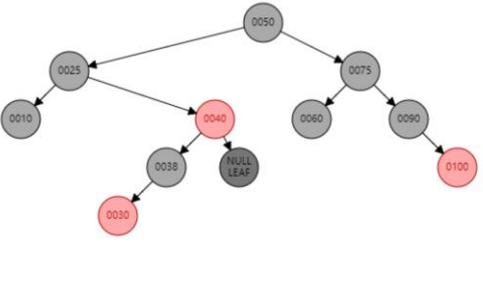 | Re-color 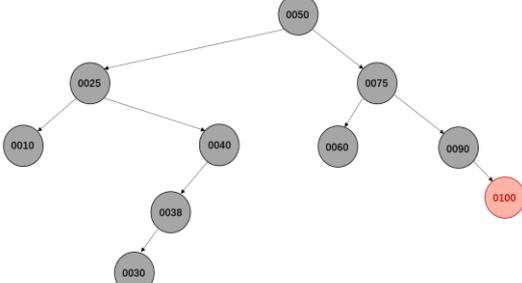 |
| 4. | Rotate 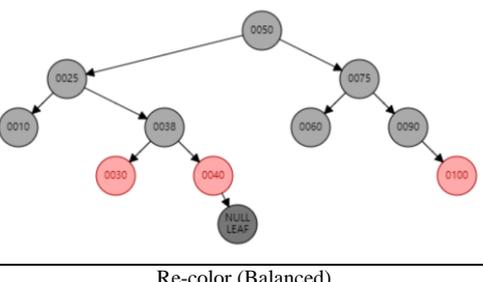 | Rotate 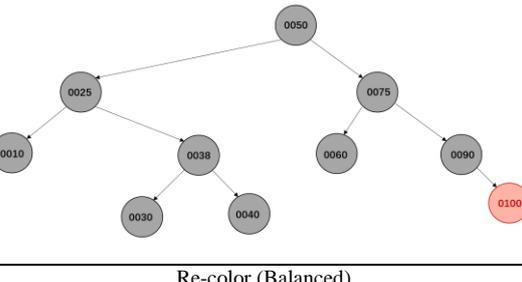 |
| 5. | Re-color (Balanced) 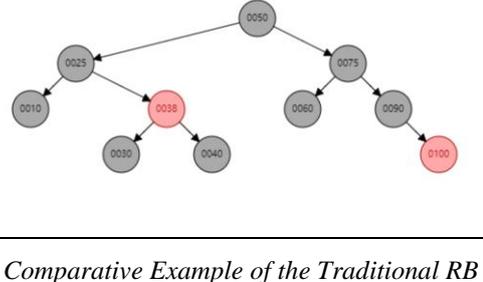 | Re-color (Balanced) 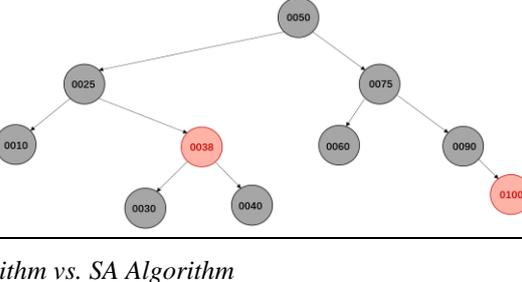 |

### 4.6.2  Second Comparative Example of the Traditional RB Algorithm vs. SA Algorithm

Table 11 is another comparison of the TA with our SA algorithm. The side-by-side pictorial evaluation as itemized by the numbering of the steps taken depicts that the SA balanced the RB tree in 4 steps as against the TA in 5 steps. That is, the SA algorithm finished one step faster than the TA RB algorithm. From the step-by-step analysis, it has been shown in this work that the SA algorithm is clearer and much more precise for teaching and learning the **DB** removal problem.

Table 11. Side-by-Side Performance Comparison Where SA Performed Faster than TA.

| Steps | Traditional Algorithm | Symbolic-Arithmetic Algorithm |
|---|---|---|
| 1. | Delete 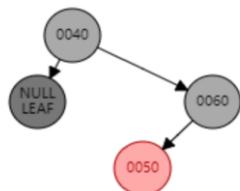 | Delete 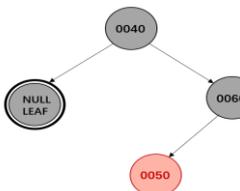 |
| 2. | Rotate 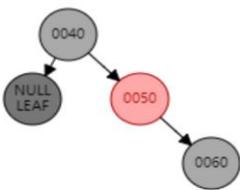 | Rotate 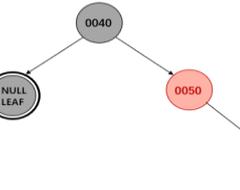 |
| 3. | Re-color 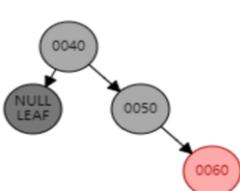 | Re-color 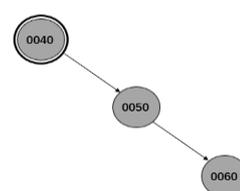 |
| 4. | Rotate 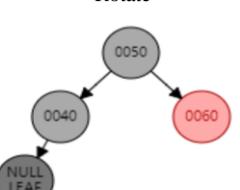 | Rotate (Balanced) 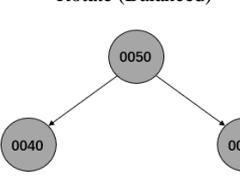 |
| 5. | Re-color (Balanced) 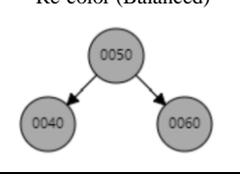 | |

Figure 19 shows a graphical representation of the TA and SA algorithm from the demostrations presented in Table 10 and 11, respectively.

| Runtime Data of the TA versus SA Algorithm for DB Removal and Tree Balancing | | |
|---|---|---|
| | R(p) and InnerR(r) | R(p) and InnerR(r) |
| Traditional Algorithm (TA) | 5 | 5 |
| Symbolic Algorithm (SA) | 5 | 4 |

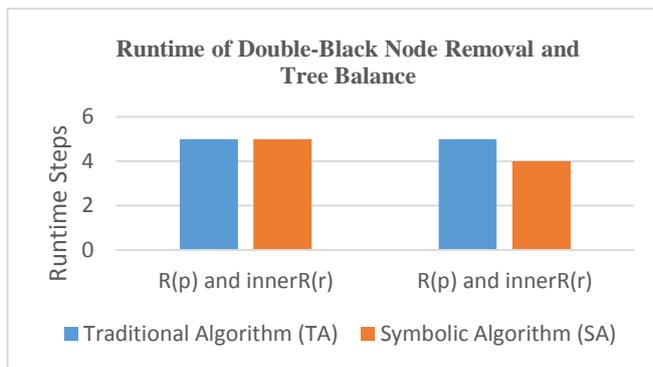

Fig. 19. Runtime visualization of the TA versus SA algorithm. On one RB tree problem, both TA and SA showed equal runtime of 5 steps. On another problem, the SA had a faster runtime of 4 steps as against the 5 steps of the TA.

### 4.6.3 Feedback from the Symbolic Algebraic Teaching Approach

As first reported in [4], for over 4 academic years running, this work on the symbolic-algebraic arithmetic (SA) method of teaching **DB** removal in RB trees has been ongoing. The questions and feedback *w.r.t.* the SA algorithmic method from students are hereby stated below; where SA represents symbolic-algebra arithmetic, and TA means the traditional (or conventional) algorithm of RB trees.

*(a) Question 1: What is your opinion on this new symbolic method in the understanding of the removal of DB and balancing of the tree*?

Table 11. Feedback on the Understanding of the Symbolic-Algebraic Method [4].

| Student Feedback | Analysis |
|---|---|
| For me, the new [symbolic] algebra (SA) algorithm is easier to understand compared to the conventional approach, with less classifications and more symbolic visual aids. But it still requires practicing to get the hang of it. | * SA approach is simpler.<br>* SA is visual and animated in process.<br>* SA has less classification.<br>* SA is needed to be learned too. |
| Symbolic algebraic application to tree balance after rotations and node coloration has a fixed pattern to follow which is lacking in the traditional algorithm. | * SA provides a fixed length of steps. |
| The symbols method provides better understanding for node coloration. It explains what node needs a color change, why it needs to change and how it will change. | * SA gives clearer understanding.<br>* SA gives the step and the exact node to color. |

*(b) Question 2: What are the perceived differences to other approaches in literature and the original DB removal algorithm?*

Table 12. Feedback on the Conventional RB Tree Algorithm vs. the Symbolic-Algebraic Method.

| Student Feedback | Analysis |
|---|---|
| This [symbolic approach] is definitely simpler for me because it simplifies the process inherent in a traditional algorithm (TA). Both the traditional red-black tree algorithm and the new symbolic approach algorithm involves remembering several conditions for deletion and rotation before carrying out corresponding operations. In this case [symbolic approach], less is definitely better, because it's not as complicated as it used to be. | * SA is definitely simpler.<br>* Both the TA and new SA have certain conditions to learn (w.r.t. rotation); but the SA is lesser with information.<br>* SA method is not complicated. |
| Traditional algorithm is somehow complicated. Not knowing where to stop color exchange between nodes sometime. The symbolic operation is a much simplified method of tree balancing after a node removal. The symbolic method uses a step-by-step mathematical approach that is certain about when to end color change. | * TA algorithm is complicated.<br>* TA is difficult to predict where color change will end.<br>* SA algorithm is much simpler.<br>* SA is step-by-step. |
| The symbols approach takes away the complications involved in the heavy information that surrounds the traditional algorithm. This is because there are few information to process as we fix the colors of nodes with the symbols method. The steps are easy to follow. | *SA is not complicated.<br>* SA has less information w.r.t. its operation. |
| The deletion operations of traditional red-black [algorithm] are badly explained in many resources I read including my textbook. These operations are not obvious to understand and confuse many students when they first see it. The new symbolic method provides steps to memorize the operations and makes removal of DB easier to learn." | *TA is not clear to understand.<br>* SA is a step-by-step operation.<br>* SA makes DB removal easier to memorize and learn. |

## 5. Conclusions and Further Work

The removal of a double-black (DB) node mechanically poses a lot of difficulty in teaching, learning, and understanding of red-black (RB) trees. The question of when does rotation occur? Or which node changes its color? Or at what point will a node change its color have all been addressed in this study. This study formulated and introduced three rules that uses symbolic arithmetic (SA) operations for removing DB nodes and balancing RB trees. The simplistic SA formulas which are namely the *generic rule*, *partial rule1*, and *partial rule2* were demonstrated with the different RB LR, RL, LL, and RR cases in this paper. With these rules, students are well-equipped to understand when to remove DB nodes, and where and what nodes need recoloring after rotation(s). The use of these equations or rules to remove DB nodes and recoloring of nodes in RB trees has proven to be decisive and unambiguous. By combining these formulas, we have shown that a DB can be easily removed and the process, operations, and the stages involved can also easily be understood. This paper also showed a side-by-side comparison of the traditional RB algorithm (TA) with our SA algorithm. Our findings revealed that the SA algorithm was never slower than the TA algorithm. Instead, the SA has proven to be faster in some cases and equal in other cases. More so, the paper has engaged the use of first-order logic statements in describing the nodes, their values, and operations. Also, these logic expressions have enhanced learning and made the process of dB removal explicitly explainable — in the concept of explainable AI. Thus, based on RB tree image visuals with specified logical formalism, intelligent system AI can learn and apply the SA algorithm and also explain the reasons behind every step that is taken and the decisions made in the reconstruction of the RB trees. From the foregoing, in future work, we shall look at image visuals for RB tree construction, DB removal, and tree rebalancing in the contest of explainable AI in combination with formal logic and the SA rules. Having shown that the SA runtime (number of steps taken) can be faster than the TA, further work will also look at the place of the SA algorithm in terms of time complexity.

## Acknowledgement

We would like to acknowledge all past and present students of Data Structures & Algorithms at the Wenzhou Kean University (WKU) for their collaboration, practice, and testing of this simplified symbolic-arithmetic algebraic method for satisfiability of the properties of the red-black tree.

**Authors' Profiles**

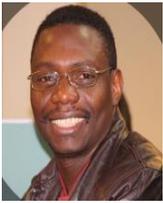

**Kennedy E. Ehimwenma,** Ph.D. Depart of Computer Science Wenzhou Kean University, China. Research interest: Multi-agent systems, knowledge representation, ontology, computational logic, IoT, smart systems, student learning, tree based learning.

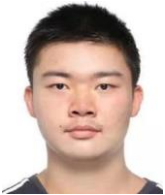

**Hongyu Zhou,** Student, Depart of Computer Science Wenzhou Kean University, China. Research interest: Blockchain, generative artificial intelligence, cryptocurrency, tree based learning.

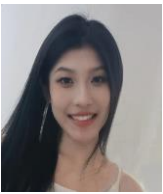

**Junfeng Wang,** Student, Depart of Computer Science Wenzhou Kean University, China. Research interest: Data structures & algorithms, machine learning, deep learning, computer graphics.

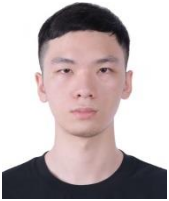

**Ze Zheng,** Student, Depart of Computer Science Wenzhou Kean University, China. Research interest: Data structures & algorithms.